\documentclass[10pt,aps,prd,twocolumn,amsmath,amssymb,nofootinbib,eqsecnum,superscriptaddress]{revtex4-2}
\usepackage{url}
\usepackage[english]{babel}
\usepackage{graphicx}
\usepackage{amsmath,bm}
\usepackage[breaklinks,colorlinks,citecolor=blue]{hyperref}
\usepackage{hhline}
\usepackage{amssymb}
\usepackage{amsthm}
\usepackage{mathrsfs}

\begin{document}
\title{Ghost-free infinite-derivative dilaton gravity in two dimensions}
\author{Ulrich K. Beckering Vinckers}
\email{bckulr002@myuct.ac.za}
\affiliation{Cosmology and Gravity Group, Department of Mathematics and Applied Mathematics,
University of Cape Town, Rondebosch 7701, Cape Town, South Africa}
\affiliation{Van Swinderen Institute, University of Groningen, 9747 AG Groningen, The Netherlands}

\author{\'Alvaro de la Cruz-Dombriz}
\email{alvaro.dombriz@usal.es}
\affiliation{Cosmology and Gravity Group, Department of Mathematics and Applied Mathematics,
University of Cape Town, Rondebosch 7701, Cape Town, South Africa}
\affiliation{ Departamento de F\'isica Fundamental, Universidad de Salamanca,
    P. de la Merced, 37008 Salamanca, Spain}
\author{\\ Ivan Kol\'a\v{r}}
\email{i.kolar@rug.nl}
\affiliation{Van Swinderen Institute, University of Groningen, 9747 AG Groningen, The Netherlands}
\author{Francisco J. Maldonado Torralba}
\email{fmaldo01@ucm.es}
\affiliation{Laboratory of Theoretical Physics, Institute of Physics, University of Tartu, W. Ostwaldi 1, 50411 Tartu, Estonia}

\author{Anupam Mazumdar}
\email{anupam.mazumdar@rug.nl}
\affiliation{Van Swinderen Institute, University of Groningen, 9747 AG Groningen, The Netherlands}
\begin{abstract}
We present the ghost-free infinite-derivative extensions of the Spherically-Reduced Gravity (SRG) and Callan-Giddings-Harvey-Strominger (CGHS) theories in two space-time dimensions. For the case of SRG, we specify the Schwarzschild-type gauge and diagonalise the quadratic action for field perturbations after taking the background fields to be those of the flat-space solution with a linear dilaton. Using the obtained diagonalisation, we construct ghost-free infinite-derivative modifications of the SRG theory. In the context of this modified SRG theory we derive a non-local modification of the linearised spherically-reduced Schwarzschild solution. For the case of CGHS gravity, we work in the conformal gauge and diagonalise the quadratic action associated with this theory for a general background solution. Using these results, we construct the ghost-free infinite-derivative modifications of the CGHS theory and examine non-local modifications to the linearised CGHS black-hole solution.
\end{abstract}
\maketitle
\section{Introduction}
It is well-known that the theory of gravity described by the Einstein-Hilbert action together with two-dimensional space-time is topological \cite{fletcher}. This is the result of the fact that, in two dimensions, the space of two-form fields is of dimension one which implies the vanishing of the Einstein tensor \cite{Ikeda:1992qz,fletcher,deLacroix:2016hpc}. Thus, one option to source gravity in two space-time dimensions is to introduce a scalar field, dubbed the dilaton field, coupled to the Ricci scalar. A number of two-dimensional theories involving the dilaton field are then possible (see \cite{Grumiller:2002nm,nojiri,Jackiw:1995hb,grumiller,Pelzer:1998ea} for a review). Among these there is the so-called Spherically-Reduced Gravity (SRG) \cite{Grumiller:2002nm,brown1988lower,Berger:1972pg,Thomi:1984na,Hajicek:1984mz,Berger:1972pg,Benguria:1976in,Grumiller:1999rz,Kastrup:1993br,Kuchar:1994zk,Lau:1995fr,Thiemann:1992jj}, 
and Callan–Giddings–Harvey–Strominger (CGHS) gravity \cite{Callan:1992rs}.

The CGHS theory, which is discussed in the seminal paper \cite{Callan:1992rs}, has also been studied extensively in \cite{Mikovic:1992id,Cangemi:1995yz,Benedict:1996qy,Mikovic:1997xm,Varadarajan:1997qz,Bilal:1992kv,Bose:1995pz,Cruz:1995zt,Hamada:1992zw,Hamada:1992hh,Kim:1999wa,Almheiri:2014cka,Ashtekar:2008jd,Ashtekar:2010qz,Grumiller:2003mc,Hayward:2001ma}. The theory admits a black-hole (BH) vacuum solution which was first presented in \cite{Witten:1991yr} and subsequently examined further in \cite{Mandal:1991tz,Elitzur:1990ubs,Dijkgraaf:1991ba,McGuigan:1991qp,Ishibashi:1991wh,DeAlwis:1991vz,Khastgir:1991ip,DeAlwis:1991vz}. 
For discussions on more general dilaton gravity theories, we direct the reader to \cite{grumiller,Russo:1992yg,Odintsov:1991qu,Grumiller:2002nm,Grumiller:2021cwg,Pelzer:1998ea,Cavaglia:1998yp}.

In the present work, we construct non-local modifications to the SRG and CGHS theories of gravity that are ghost-free. One can introduce non-locality into a given theory by including infinitely many covariant derivatives in the action \cite{SravanKumar:2019eqt,Krasnikov:1987yj,Biswas:2005qr,Biswas:2011ar,Buoninfante:2018mre,Frolov:2015bta}. In particular, one can construct ghost-free infinite-derivative modifications of General Relativity; often called Infinite Derivative Gravity (IDG) \cite{Deser:2007jk,Woodard:2014iga,Belgacem:2017cqo,Biswas:2010zk,Biswas:2012bp,Biswas:2013cha,Biswas:2016egy,Tsamis:2014hra,Buoninfante:2018xiw,Conroy:2014eja,Dimitrijevic:2020dzo}. As discussed in \cite{Dragovich:2017kge,Dragovich:2009hd}, there is motivation from $p$-adic string theory for the introduction of non-local operators and the first such application of $p$-adic mathematics in string theory appeared when studying scalar tachyon strings. In addition, it was noted in \cite{Modesto:2011kw,Modesto:2014lga,Tomboulis:2015esa,Tomboulis:1997gg,Modesto:2017hzl,Buoninfante:2018mre} that certain quantum gravity models are renormalisable through the inclusion of non-local operators. The initial-value problem in the context of infinite-derivative theories has been studied in \cite{Barnaby:2007ve,Barnaby:2008tc,Barnaby:2010kx,Gorka:2012hs,Calcagni:2018lyd,Moeller:2002vx} and the Hamiltonian formulation for non-local theories is discussed in \cite{llosa1994hamiltonian,Gomis:2000gy,Gomis:2003xv,Kolar:2020ezu,Heredia:2021wja,Heredia:2022mls}. Cosmological implications, such as the resolution of cosmological singularities, of IDG are discussed in \cite{Biswas:2006bs,Biswas:2005qr,Koshelev:2017tvv,SravanKumar:2018dlo,Biswas:2010zk,Biswas:2012bp,Koshelev:2018rau}. Studies regarding the resolution of BH singularities through the inclusion of infinitely many derivatives in the action have been conducted in \cite{Biswas:2011ar,Koshelev:2018hpt,Frolov:2015usa,Frolov:2015bia}. In the present work, we shall study the effect of non-locality in the linearised regime. Linearised solutions in the context of IDG have been studied in \cite{Biswas:2011ar,Modesto:2010uh,Frolov:2015bta,Boos:2018bxf,Kolar:2020bpo,Frolov:2016xhq,Kolar:2021oba} and we employ some of these methods in the present work. While we restrict ourselves to the linearised regime, we note that exact solutions in the context of IDG have been found in \cite{Kilicarslan:2019njc,Dengiz:2020xbu,Kolar:2021rfl,Kolar:2021uiu} whereas in two-dimensional gravity, ghost-free infinite-derivative modifications of the Polyakov action have been studied in \cite{Boos:2019vcz}.

This communication is organised as follows: In Section \ref{sec:local_theories}, we begin by considering a generalised dilaton action, defined on a two-dimensional space-time, which generates the SRG and CGHS theories of gravity. There, we shall briefly review this generalised dilaton model and derive the corresponding quadratic action without fixing the gauge. In order to diagonalise such a quadratic action, we shall consider the SRG and CGHS gravity theories separately. Thus, we shall first study the SRG theory in the Schwarzschild-type gauge in Section \ref{sec:SRG}. Herein, after specifying that gauge, we shall diagonalise the quadratic action in order to construct ghost-free infinite-derivative modifications to the SRG theory. We then construct a source term that can be used to generate the linearised Schwarzschild solution in the context of the local diagonalised theory. Using the same source in the diagonalised non-local theory, we obtain a non-local modification to the linearised spherically-reduced Schwarzschild solution. We will show that the singular nature of the linearised local solution is resolved in the non-local theory through the appearance of the error function. This is comparable to the linearised non-local Schwarzschild solution of IDG obtained in \cite{Biswas:2011ar} where the singular nature is also resolved by the error function.

In Section \ref{sec:IDG_CGHS}, we shall turn our attention to the CGHS theory. In this case, we shall resort to the conformal gauge and review the well-known CGHS BH solution in this gauge. The corresponding quadratic action is then diagonalised for a general background solution in order to construct ghost-free infinite-derivative modifications to the CGHS theory. Accordingly we shall introduce a source term that generates the CGHS BH solution of the local theory in the linearised regime. Using the aforementioned source term together with the non-local theory, we shall examine how the linearised solution is modified as a result of introducing non-locality. We will show that the singular nature appearing in the linearised local solution is resolved in the non-local theory through the appearance of the complementary error function.

We shall end up with Section \ref{Conclusions} which encapsulates the conclusions of this investigation. 

\section{Local two-dimensional dilaton gravity}\label{sec:local_theories}
\subsection{Action and quadratic action}
Let us consider the following dilaton action in two-dimensional space-time
\begin{align}\label{eq:action_with_phi}
S_{\text{local}}:=4\int\textup{d}^2x\sqrt{-g}\bigg[\frac{R}{4}+k(\partial\phi)^2
+\lambda^2+\frac{a^2}{2}{\text e}^{2\phi}\bigg]{\text e}^{-2\phi}\ ,
\end{align}
where $g_{\mu\nu}$ is the metric tensor, $R$ is the Ricci scalar, $\phi$ is the dilaton field and $k,\lambda$ and $a$ are constants \cite{Grumiller:2002nm,grumiller}. We note that both $\lambda$ and $a$ have dimensions of length$\phantom{}^{-1}$ while $k$ is dimensionless. By defining the field $\Phi:={\text e}^{-\phi}$, this action \eqref{eq:action_with_phi} can be written as
\begin{align}\label{eq:action_with_Phi}
S_{\text{local}}=4\int\text{d}^2x\sqrt{-g}\bigg[\frac{\Phi^2R}{4}+&k\left(\partial\Phi\right)^2
+\Phi^2\lambda^2+\frac{a^2}{2}\bigg]\ .
\end{align}
This generalised action describes a number of dilaton gravity theories. In particular, by setting $\lambda=0$, $k=1/2$ and leaving $a$ unspecified, the action \eqref{eq:action_with_Phi} is that of SRG \cite{Grumiller:2002nm,brown1988lower,Berger:1972pg,Thomi:1984na,Hajicek:1984mz}. The case where $k=1$, $a=0$ and leaving $\lambda$ unspecified corresponds to the vacuum CGHS theory \cite{Callan:1992rs}. Table \ref{table:dilaton_model_parameters} lists these two theories and their corresponding parameter specifications. Although not examined in this communication, the case where $a=0$, $k=0$ and leaving $\lambda$ unspecified is the so-called Jackiw-Teitelboim gravity theory \cite{Jackiw:1984je,Teitelboim:1983ux}. In addition, there are more general dilaton gravity actions which are discussed in \cite{Grumiller:2002nm}, however, in this paper the only dilaton gravity models that we consider are the SRG and CGHS theories.
\begin{center}
\setlength{\tabcolsep}{12pt}
\begin{table}
\begin{tabular}{| c | c | c | c |}
\hline
	  & $k$ & $\lambda$ & $a$ \\
	  \hline
 SRG & $1/2$ & $0$ & $a$ \\ 
 \hline
 CGHS & $1$ & $\lambda$ & $0$ \\ 
 \hline
\end{tabular}
\caption{Parameter specifications that, upon substitution into the action \eqref{eq:action_with_Phi}, generate either the SRG or CGHS theories.}\label{table:dilaton_model_parameters}
\end{table}
\end{center}

In order to state the definition for the quadratic action, we first perform perturbations of the metric and dilaton field as
\begin{align}
g_{\mu\nu}=\bar g_{\mu\nu}+\delta g_{\mu\nu}\ ,\ \ \ \ \ \ \ \ \ \ \Phi=\bar\Phi+\delta\Phi\ ,
\end{align}
where $\left(\bar g_{\mu\nu},\bar\Phi\right)$ is a solution to the equations of motion and $\delta g_{\mu\nu}$ and $\delta\Phi$ are the perturbed metric and perturbed dilaton field respectively. We also require that the perturbed dilaton field satisfies
\begin{align}\label{eq:smallness_general}
|\delta\Phi|\ll|\bar\Phi|\ ,
\end{align} 
and that the curvature scale of the metric perturbations be much smaller than that of the background metric in order for our approximations to be valid. These conditions placed on the perturbed dilaton are referred to as the \text{smallness conditions}. In this paper, we make use of the following definition for the \textit{quadratic action}
\begin{align}\label{eq:quadratic_action_definition}
&\delta^2S_{\text{local}}:=\int\text{d}^2x\ \text{d}^2x'\Bigg[\delta\Phi(x)\delta\Phi(x')\frac{\delta^2S_{\text{local}}}{\delta\Phi(x)\delta\Phi(x')}\Bigg|_{\left(\bar g_{\mu\nu},\bar\Phi\right)}\nonumber \\
&+\delta\Phi(x)\delta g^{\mu\nu}(x')\frac{\delta^2S_{\text{local}}}{\delta\Phi(x)\delta g^{\mu\nu}(x')}\Bigg|_{\left(\bar g_{\mu\nu},\bar\Phi\right)}\nonumber\\
&+\delta g^{\mu\nu}(x)\delta\Phi(x')\frac{\delta^2S_{\text{local}}}{\delta g^{\mu\nu}(x)\delta\Phi(x')}\Bigg|_{\left(\bar g_{\mu\nu},\bar\Phi\right)}\nonumber\\
&+\delta g^{\mu\nu}(x)\delta g^{\alpha\beta}(x')\frac{\delta^2S_{\text{local}}}{\delta g^{\mu\nu}(x)\delta g^{\alpha\beta}(x')}\Bigg|_{\left(\bar g_{\mu\nu},\bar\Phi\right)}\Bigg]\ ,
\end{align}
whose variation with respect to the fields gives the field equations for the perturbations at first order.

Using the previous definition, we now wish to derive the quadratic action associated with \eqref{eq:action_with_Phi}. Indeed, through the variation of \eqref{eq:action_with_Phi}, one can obtain
\begin{align}\label{eq:first_variation}
&\delta S_{\text{local}}=-2\int\text{d}^2x\sqrt{-g}g_{\mu\nu}\delta g^{\mu\nu}\bigg[\frac14\Phi^2R+k\left(\partial\Phi\right)^2\nonumber\\
&+\Phi^2\lambda^2+\frac{a^2}{2}\bigg]+4\int\text{d}^2x\sqrt{-g}\bigg[\frac12\Phi\delta\Phi R\nonumber\\
&+
\frac{\Phi^2}{4}\Big(R_{\mu\nu}\delta g^{\mu\nu}-\nabla_\mu \nabla_\nu\delta g^{\mu\nu}+g_{\mu\nu}\Box\delta g^{\mu\nu}\Big)\nonumber\\
&+k\delta g^{\mu\nu}\partial_\mu\Phi\partial_\nu\Phi+2k\partial^\nu\Phi\partial_\nu\delta\Phi
+2\Phi\delta\Phi\lambda^2\bigg]\ .
\end{align}
This expression \eqref{eq:first_variation} would allow us to compute the first-order functional derivatives of the action. The functional derivative with respect to the dilaton field $\Phi$ is
\begin{align}\label{eq:first_variation_dilaton}
\frac{\delta S_{\text{local}}}{\delta\Phi}=4\sqrt{-g}\left[\frac12\Phi R-2k\Box\Phi+2\Phi\lambda^2\right]\ ,
\end{align}
whereas the functional derivative of the action with respect to the inverse metric $g^{\mu\nu}$ is
\begin{align}\label{eq:first_variation_metric}
&\frac{\delta S_{\text{local}}}{\delta g^{\mu\nu}}=-2\sqrt{-g}\ g_{\mu\nu}\left[k\left(\partial\Phi\right)^2+\Phi^2\lambda^2+\frac{a^2}{2}\right]\nonumber\\
&+\sqrt{-g}\left[g_{\mu\nu}\Box\Phi^2-\nabla_\mu \nabla_\nu\Phi^2+4k\partial_\mu\Phi\partial_\nu\Phi\right]\ ,
\end{align}
where in the three previous expressions $\Box:=g^{\mu\nu}\nabla_\mu \nabla_\nu$  is the usual space-time covariant d'Alembertian operator and $\nabla_\mu$ is the Levi-Civita covariant derivative. Also note that in obtaining \eqref{eq:first_variation_metric} we have made use of the fact that the Einstein tensor vanishes identically in two dimensions, i.e., $R_{\mu\nu}=\frac12Rg_{\mu\nu}$. This results in the Ricci tensor not appearing in \eqref{eq:first_variation_metric}. It is worth mentioning that the trace of \eqref{eq:first_variation_metric} renders
\begin{align}\label{eq:first_variation_metric_trace}
g^{\mu\nu}\frac{\delta S_{\text{local}}}{\delta g^{\mu\nu}}=\sqrt{-g}\left[\Box\Phi^2-4\left(\Phi^2\lambda^2+\frac{a^2}{2}\right)\right]\ .
\end{align}
Thus, the relevant equations of motion are obtained by setting expressions \eqref{eq:first_variation_dilaton} and \eqref{eq:first_variation_metric} equal to zero.

On the other hand, by computing the second-order functional derivatives associated with the action \eqref{eq:action_with_Phi}, we can use the definition \eqref{eq:quadratic_action_definition} in order to arrive at the quadratic action
\begin{align}\label{eq:quadratic_action_general}
&\delta^2S_{\text{local}}=\int\text{d}^2x\sqrt{-\bar g}\bigg\{8k\delta\Phi\left(\frac{\bar\Box\bar\Phi}{\bar\Phi}-\bar\Box\right)\delta\Phi\nonumber\\
&+4\delta g^{\mu\nu}\bigg[\frac12\bar R\bar g_{\mu\nu}\bar\Phi\delta\Phi+\Big(\bar g_{\mu\nu}\bar\Box-\bar \nabla_\mu\bar \nabla_\nu\Big)\left(\bar\Phi\delta\Phi\right)\nonumber\\
&+4k\partial_{\mu}\bar\Phi\partial_{\nu}\delta\Phi-(1+k)\bar g_{\mu\nu}\bar \nabla_\alpha\left(\delta\Phi\partial^\alpha\bar\Phi\right)\nonumber\\
&+(1-k)\bar g_{\mu\nu}\delta\Phi\bar\Box\bar\Phi
\bigg]\nonumber\\
&+\delta g^{\mu\nu}\bigg[\bar g_{\mu\alpha}\Big(4k\partial_\beta\bar\Phi\partial_\nu\bar\Phi-\bar \nabla_\nu\bar \nabla_\beta\bar\Phi^2\Big)\nonumber\\
&+\bar g_{\mu\nu}\bar \nabla_\alpha\bar\Phi^2\bar \nabla_\beta-\bar g_{\mu[\nu}\bar g_{\alpha]\beta}\bar \nabla^\sigma\bar\Phi^2\bar \nabla_\sigma\nonumber\\
&-\bar g_{\beta\mu}\bar \nabla_\alpha\bar\Phi^2\bar \nabla_{\nu}
\bigg]\delta g^{\alpha\beta}\bigg\}\ ,
\end{align}
whose derivation can be found in Appendix \ref{sec:derivation_quadratic} where the functional derivatives \eqref{eq:first_variation_dilaton} and \eqref{eq:first_variation_metric} are used.

Equation \eqref{eq:quadratic_action_general} is the quadratic action associated with the general dilaton model \eqref{eq:action_with_Phi} without any gauge fixing. In addition, this quadratic action is valid for a general background solution $\left(\bar g_{\mu\nu},\bar\Phi\right)$ provided that the smallness conditions stated in \eqref{eq:smallness_general} are satisfied. In Sections 
\ref{sec:SRG}
and \ref{sec:IDG_CGHS}
 we shall consider the SRG and CGHS theories independently. For each of those two theories, we shall diagonalise the quadratic action after a gauge fixing. Before proceeding, we wish to make a note on the gauge fixing of the action \eqref{eq:action_with_Phi}. In this work, we will either specify the so-called conformal gauge or the Schwarzschild-type gauge \cite{Grumiller:2002nm} which are defined in later sections. In both cases, the gauge-fixed action contains all the dynamics of the original action \eqref{eq:action_with_Phi}. That is, one obtains the same dynamical equations of motion regardless of whether one fixes the gauge at the level of the action or at the level of the field equations. The constraint equations, on the other hand, appear as a result of specifying a gauge. In Appendices \ref{sec:dilaton_gravity_in_conformal_gauge} and \ref{sec:dilaton_gravity_in_Schwarzschild_gauge} we show this explicitly for both choices of gauges. For a more elaborate treatment of gauge fixing in the context of two-dimensional dilaton gravity models, the interested reader is directed to \cite{Kummer:1996hy,Giddings:1992ae}.

\subsection{Symmetries of the dilaton action}
\label{sec:degrees_of_freedom}
Let us briefly examine the redundancies present in the gravitational theories generated by \eqref{eq:action_with_Phi} (for a detailed discussion, the interested reader is directed to \cite{deLacroix:2016hpc,Cruz:1996ze,fletcher}). The metric tensor and dilaton field contribute three and one degrees of freedom respectively. The action \eqref{eq:action_with_Phi} admits diffeomorphism invariance contributing two redundancies. Given these two redundancies, after a gauge fixing, one can write the action \eqref{eq:action_with_Phi} as a functional of two scalar fields: one scalar field describing the dilaton and the other describing the metric. In the case of the SRG theory, i.e., when we set $\lambda=0$ and $k=1/2$, this leaves us with two propagating scalar fields. In the case of the CGHS theory, i.e., when $a=0$ and $k=1$, there is one additional symmetry described by the following transformation
\begin{align}\label{eq:CGH_additional_symmetry}
\delta g^{\mu\nu}=\frac{2\varepsilon g^{\mu\nu}}{\Phi^2}\ ,\ \ \ \ \ \ \ \ \ \ \delta\Phi=\frac{\varepsilon}{\Phi}\ ,
\end{align}
where $\varepsilon\in\mathbb{R}$ is a constant. We will return to this symmetry in Section \ref{sec:IDG_CGHS} where the CGHS theory is discussed separately.

\subsection{Source action }
When studying the SRG and CGHS theories in the linearised regime, we will introduce some source action in addition to the geometric, either local or non-local, action. The purpose of introducing such a source action is to generate solutions for the local theories that satisfy the following two properties:
\begin{enumerate}
\item \label{itm:condition_1} the local solutions are singular when considering the entire space-time;
\item \label{itm:condition_2} the local solutions coincide with the BH solution of the relevant theory in the space-time region for which the smallness conditions are satisfied.
\end{enumerate}

Once we have identified a source action accomplishing the two properties above in the local theory, we will make use of the same source action when studying the non-local theory. 
In particular, we shall be interested in examining how the singular nature is resolved as a result of non-locality as well as how the local BH solutions of the SRG and CGHS theories are modified in the region for which the smallness conditions are satisfied. A similar analysis is done for the case of four-dimensional IDG in  \cite{Biswas:2011ar} where a non-local modification to the linearised Schwarzschild solution of GR is obtained.

For our purposes, we consider a source action of the form
\begin{align}\label{eq:general_source_action_sp}
S_{\text{source}}=4\int\text{d}^2x\sqrt{-g}\ U(\Phi,\bar g_{\mu\nu},\bar\Phi)\ ,
\end{align}
where $U(\Phi,\bar g_{\mu\nu},\bar\Phi)$ is some function of the dilaton field $\Phi$ and the background solution $(\bar g_{\mu\nu},\bar\Phi)$ to the field equations when $U=0$. Thus the total action under consideration would be
\begin{align}\label{eq:total_action_general_no_gauge}
S_{\text{total}}:=S_{\text{local}}+S_{\text{source}}\ ,
\end{align}
where the local action $S_{\text{local}}$ and the source action $S_{\text{source}}$ are given by \eqref{eq:action_with_Phi} and \eqref{eq:general_source_action_sp} respectively. From the definition given in \cite{Mann:1992yv}, the stress-energy tensor is
\begin{align}\label{eq:stress_energy_tensor}
&T_{\mu\nu}:=\frac{1}{\sqrt{-g}}\frac{\delta S_{\text{total}}}{\delta g^{\mu\nu}}\nonumber\\
&=-2g_{\mu\nu}\left[k\left(\partial\Phi\right)^2+\Phi^2\lambda^2+\frac{a^2}{2}+U(\Phi,\bar g_{\mu\nu},\bar\Phi)\right]\nonumber\\
&+g_{\mu\nu}\Box\Phi^2-\nabla_\mu \nabla_\nu\Phi^2+4k\partial_\mu\Phi\partial_\nu\Phi\ ,
\end{align}
where we made use of equation \eqref{eq:first_variation_metric}. In addition, by making use of equation \eqref{eq:first_variation_dilaton} and varying the source action with respect to $\Phi$ one can obtain
\begin{align}\label{eq:first_variation_dilaton_with_V}
\frac{\delta S_{\text{total}}}{\delta\Phi}=4\sqrt{-g}\bigg[\frac{\Phi R}{2}& -2k\Box\Phi
\nonumber\\
&+2\Phi\lambda^2+\frac{\partial U(\Phi,\bar g_{\mu\nu},\bar\Phi)}{\partial\Phi}\bigg]\ .
\end{align}
To examine whether the stress-energy tensor \eqref{eq:stress_energy_tensor} is conserved, we take the divergence and write
\begin{align}
&\nabla^\mu T_{\mu\nu}=-\Box \nabla_\nu\Phi^2+\nabla_\nu\Box\Phi^2\nonumber\\
&-2\left[2\Phi\lambda^2+\frac{\partial U(\Phi,\bar g_{\mu\nu},\bar\Phi)}{\partial\Phi}+2k\Box\Phi\right]\partial_\nu\Phi\ .
\end{align}
From the definition of the Riemann tensor, the first two terms on the right-hand-side can be written as
\begin{align}
-\Box \nabla_\nu\Phi^2+\nabla_\nu\Box\Phi^2=-R_\nu^{\ \mu}\nabla_\mu\Phi^2\ .
\end{align}
Since the Einstein tensor is identically zero in two space-time dimensions, i.e., $R_{\mu\nu}=\frac12Rg_{\mu\nu}$, we have
\begin{align}
\nabla^\mu T_{\mu\nu}=-2\bigg[\frac12R&\Phi+2\Phi\lambda^2\nonumber\\
&+\frac{\partial U(\Phi,\bar g_{\mu\nu},\bar\Phi)}{\partial\Phi}-2k\Box\Phi\bigg]\partial_\nu\Phi\ .
\end{align}
The right-hand-side of the last expression can be identified with the right-hand-side of equation \eqref{eq:first_variation_dilaton_with_V} which implies that the divergence of the stress-energy tensor is
\begin{align}
\nabla^\mu T_{\mu\nu}=-\frac{\partial_\nu\Phi}{2\sqrt{-g}}\frac{\delta S_{\text{total}}}{\delta\Phi}\ .
\end{align}
As mentioned in \cite{Mann:1992yv}, this implies that the stress-energy tensor is conserved whenever the dilaton equation of motion associated with $S_{\text{total}}$ is satisfied. Although additional properties of the stress-energy tensor are studied in \cite{Mann:1992yv}, we do not discuss the topic further here and instead direct the interested reader to the aforesaid reference.

In this work, we are interested in considering potentials $U(\Phi,\bar g_{\mu\nu},\bar\Phi)$ of the form
\begin{align}\label{eq:our_potential_U}
U(\Phi,\bar g_{\mu\nu},\bar\Phi)=-\frac{\Phi^2\bar g^{\mu\nu}A_{\mu\nu}}{2\left(a\bar\Phi-4\lambda\bar\Phi^2\right)}\ ,
\end{align}
where $A_{\mu\nu}$ contains a Dirac delta function. Our motivation for taking the potential $U(\Phi,\bar g_{\mu\nu},\bar\Phi)$ to be of the form given in \eqref{eq:our_potential_U} is that, for an appropriate choice of $A_{\mu\nu}$, this potential satisfies both properties \ref{itm:condition_1} and \ref{itm:condition_2} for SRG and CGHS gravity. We will show this explicitly in Sections \ref{sec:SRG}  and \ref{sec:IDG_CGHS} for the SRG and CGHS theories respectively. When generating linearised solutions in these theories, we will be interested in obtaining static solutions perturbed around a flat space-time. To this end, we will take the background metric to be the Minkowski metric in Cartesian coordinates $(t,r)$, i.e., $\bar g_{\mu\nu}=\eta_{\mu\nu}:=\text{diag}(-1,1)$. In this context, we will take $A_{\mu\nu}$ to be of the form
\begin{align}\label{eq:def_A_mu_nu}
A_{\mu\nu}=M\left(\eta_{\mu\nu}+\delta_{\mu\nu}\right)\delta'(r-b)\ ,
\end{align}
where $\delta_{\mu\nu}:=\text{diag}(1,1)$, $M\in\mathbb{R}$ is a constant of dimension length$\phantom{}^{-1}$, $b\in\mathbb{R}$ is a constant describing the position of the source and the prime $'$ denotes differentiation with respect to $r$. We will show that this choice for $A_{\mu\nu}$ can be used to generate the linearised BH solutions of the local SRG and CGHS theories with the parameter $M$ coinciding with the BH mass.

As a concluding remark about the source action defined here, we wish to examine whether the latter is invariant under the symmetry transformation \eqref{eq:CGH_additional_symmetry}. While equation \eqref{eq:general_source_action_sp} yields a stress-energy tensor that is conserved when the dilaton equation of motion is satisfied, there is no guarantee that the transformation \eqref{eq:CGH_additional_symmetry}, which is a symmetry of the CGHS theory, will leave the source action invariant for a general $U(\Phi,\bar g_{\mu\nu},\bar\Phi)$. However, for the specific case where the potential is of the form \eqref{eq:our_potential_U}, the source action is invariant under the transformation \eqref{eq:CGH_additional_symmetry}. This will be shown in Section \ref{sec:IDG_CGHS} when we consider the CGHS theory independently.

Having now stated the source action to be used, we wish to examine the total action \eqref{eq:total_action_general_no_gauge} at quadratic order. By expanding the total action to quadratic order in both $\delta g_{\mu\nu}$ and $\delta\Phi$ around the $U=0$ background solution $(\bar g_{\mu\nu},\bar\Phi)$, one has
\begin{align}\label{eq:functional_taylor}
&S_{\text{total}}\approx S_{\text{local}}[\bar g_{\mu\nu},\bar\Phi]+S_{\text{source}}[\bar g_{\mu\nu},\bar\Phi]+\delta S_{\text{local}}[\bar g_{\mu\nu},\bar\Phi]\nonumber\\
&+\delta S_{\text{source}}[\bar g_{\mu\nu},\bar\Phi,\delta g^{\mu\nu},\delta\Phi]+\frac12\delta^2S_{\text{local}}[\bar g_{\mu\nu},\bar\Phi,\delta g^{\mu\nu},\delta\Phi]\nonumber\\
&+\frac12\delta^2S_{\text{source}}[\bar g_{\mu\nu},\bar\Phi,\delta g^{\mu\nu},\delta\Phi]\ .
\end{align}
The first two terms on the right-hand-side of \eqref{eq:functional_taylor} are constant with respect to functional differentiation while the third term is zero since $\bar g_{\mu\nu}$ and $\bar\Phi$ satisfy the vacuum equations of motion. In addition, the source action is of the order of the perturbations, i.e., $S_{\text{source}}\sim\mathcal{O}(\delta g_{\mu\nu})+\mathcal{O}(\delta\Phi)$. Therefore, at quadratic order, we can safely ignore the last term on the right-hand-side of \eqref{eq:functional_taylor}. It follows that the quadratic part of the total action is
\begin{align}\label{eq:functional_taylor_2}
\delta^2S_{\text{total}}&:=\delta S_{\text{source}}[\bar g_{\mu\nu},\bar\Phi,\delta g^{\mu\nu},\delta\Phi]\nonumber\\
&+\frac12\delta^2S_{\text{local}}[\bar g_{\mu\nu},\bar\Phi,\delta g^{\mu\nu},\delta\Phi]\ .
\end{align}
It is important to note that the expansion in \eqref{eq:functional_taylor} is valid provided that the smallness condition \eqref{eq:smallness_general} is satisfied.

\section{SRG gravity}\label{sec:SRG}
In this section, we consider the SRG theory which is described by the action \eqref{eq:action_with_Phi} with $k=1/2$, $\lambda=0$ and $a$ left unspecified. As already mentioned, there are two redundancies present in equation \eqref{eq:action_with_Phi} as a result of diffeomorphism invariance. We can remove these two redundancies by specifying the Schwarzschild-type gauge. In this choice of gauge, a generic metric can be written in the form \cite{Grumiller:2002nm}
\begin{align}\label{eq:Schwarzschild_gauge}
\text{d}s^2=-f(r,t)\text{d}t^2+\frac{\text{d}r^2}{f(r,t)}\ .
\end{align}
The SRG theory, which is obtained through the spherical reduction of four-dimensional GR, admits the solution described by
\begin{align}\label{eq:SRG_linearised_Schwarzschild_local_xi}
f=1-\frac{2M}{a^2r}\ ,
\end{align}
whereas the dilaton field is given by
\begin{align}\label{eq:SRG_linearised_Schwarzschild_local_dilaton}
\Phi=ar\ .
\end{align}
The parameter $M$ in equation \eqref{eq:SRG_linearised_Schwarzschild_local_xi} is the ADM mass\footnote{When performing the spherical reduction of the Einstein-Hilbert action of four-dimensional GR, the resulting action is given by \eqref{eq:action_with_Phi} with $k=1/2$, $\lambda=0$ and a prefactor proportional to $1/a^2$ which we have dropped. When taking this prefactor into account, the ADM mass is then rescaled to $M/a^2$ which has dimensions of length and coincides with the mass of the Schwarzschild solution of GR \cite{Grumiller:2002nm}.} and has dimensions of length$^{-1}$. In this note, we refer to this solution as the \textit{spherically-reduced Schwarzschild solution} \cite{Grumiller:2002nm}. It is important to note that the flat-space solution, i.e., when $M=0$, corresponds to the Minkowski metric with $f=1$ while the dilaton remains $\Phi=ar$. The modification of the Schwarzschild solution in the context of IDG is well known and has been obtained in the linearised regime in \cite{Biswas:2011ar}. Here we construct a ghost-free infinite-derivative modification of SRG and examine how the linearised spherically-reduced Schwarzschild solution is modified. In order to construct a ghost-free infinite-derivative modification of SRG, we first have to diagonalise the local quadratic action \eqref{eq:quadratic_action_general}.

\subsection{Local SRG gravity}
After specifying the Schwarzschild-type gauge \eqref{eq:Schwarzschild_gauge}, we can perturb the metric around the Minkowski solution by writing $
f=1+\delta f
$ and expanding the metric to first order in $\delta f$. In the perturbation, the background metric is $\bar g_{\mu\nu}=\eta_{\mu\nu}$ while the perturbed metric is $\delta g^{\mu\nu}=\delta^{\mu\nu}\delta f$. In addition, we take the background dilaton field to be $\bar\Phi=ar$. With these specifications, the quadratic action \eqref{eq:quadratic_action_general} simplifies considerably to
\begin{align}
\label{eq:quadratic_action_SRG}
\delta^2S^{\text{SRG}}_{\text{local}}=4\int\text{d}^2x\left(\partial_\mu\delta\Phi\partial^\mu\delta\Phi-\delta^{\mu\nu}\delta f\bar\Phi\partial_\mu\partial_\nu\delta\Phi\right)\ ,
\end{align}
where we lower and raise indices using the Minkowski metric $\eta_{\mu\nu}$. For a derivation of \eqref{eq:quadratic_action_SRG}, the interested reader is directed to Appendix \ref{sec:derivation_SRG}. While the quadratic action \eqref{eq:quadratic_action_SRG} contains all the dynamics, we also have the Schwarzschild-type gauge constraint equations
\begin{align}\label{eq:schwarzschild_type_gauge_constraint_1_v2}
f\partial^2_r\Phi^2-\frac{1}{f}\partial_t^2\Phi^2+f'\partial_r\Phi^2+\frac{\dot f\partial_t\Phi^2}{f^2}-2a^2=0\ ,
\end{align}
and
\begin{align}\label{eq:schwarzschild_type_gauge_constraint_2_v2}
\partial_t\partial_r\Phi^2-\frac{f'\partial_t\Phi^2}{2f}+\frac{\dot f\partial_r\Phi^2}{2f}-2\dot\Phi\Phi'=0\ .
\end{align}
For a derivation of these constraint equations, the interested reader is directed to Appendix \ref{sec:dilaton_gravity_in_Schwarzschild_gauge}.

In order to diagonalise the quadratic action \eqref{eq:quadratic_action_SRG}, we can factorise the integrand and obtain
\begin{align}\label{eq:quadratic_action_SRG_factorise}
\delta^2S^{\text{SRG}}_{\text{local}}&=4\int\text{d}^2x\bigg\{\left[\partial^\mu\delta\Phi+\frac12\delta^{\mu\nu}\partial_\nu\left(\bar\Phi\delta f\right)\right]^2\nonumber\\
&+\frac{1}{4}\bar\Phi\delta f\bar\Box\left(\bar\Phi\delta f\right)\bigg\}\ ,
\end{align}
where the background d'Alembertian is now $\bar\Box=\eta^{\mu\nu}\partial_\mu\partial_\nu$. Given this form of the quadratic action, it is convenient to carry out a redefinition of fields with
\begin{align}\label{eq:SRG_redefinition of fields_N}
\delta N:=\frac12\bar\Phi\delta f\ ,
\end{align}
and
\begin{align}\label{eq:SRG_redefinition of fields_V}
\delta V^{\mu\nu}:=\eta^{\mu\nu}\delta\Phi+\delta^{\mu\nu}\delta N\ .
\end{align}
In terms of these redefined fields, \eqref{eq:quadratic_action_SRG_factorise} can be written as
\begin{align}\label{eq:quadratic_action_SRG_diagonalised}
\delta^2S^{\text{SRG}}_{\text{local}}=4\int\text{d}^2x\left(\delta N\bar\Box\delta N-\delta V^\nu\phantom{}_\mu\partial_\nu\partial_\alpha \delta V^{\mu\alpha}\right)\ ,
\end{align}
which corresponds to the diagonalised quadratic action associated with SRG in the Schwarzschild-type gauge with the background solution taken to be $(\bar g_{\mu\nu},\bar\Phi)=(\eta_{\mu\nu},ar)$.

Having derived the diagonalised quadratic action for the SRG theory, we turn our attention to solving the linearised theory, provided the source action and $U(\Phi,\bar g_{\mu\nu},\bar\Phi)$ are given by \eqref{eq:general_source_action_sp} and \eqref{eq:our_potential_U} respectively. To accomplish this, let us examine \eqref{eq:general_source_action_sp} in the Schwarzschild-type gauge with $\bar g^{\mu\nu}=\eta^{\mu\nu}$. Under such a consideration, expression \eqref{eq:general_source_action_sp} yields
\begin{align}
S^{\text{SRG}}_{\text{source}}=-2\int\text{d}^2x\frac{\Phi^2\eta^{\mu\nu}A_{\mu\nu}}{a\bar\Phi}\ .
\end{align}
We remind the reader that, by definition, we set $\lambda=0$ in the SRG theory. By varying the source action above and evaluating the result at the background fields we find
\begin{align}
\delta S^{\text{SRG}}_{\text{source}}=-\frac4a\int\text{d}^2x\ \delta\Phi\eta^{\mu\nu}A_{\mu\nu}\ ,
\end{align}
which, in terms of the redefined fields $\delta N$ and $\delta V^{\mu\nu}$, becomes
\begin{align}
\label{eq:SRG_source_action}
\delta S^{\text{SRG}}_{\text{source}}=-\frac4a\int\text{d}^2x\ \left(\delta V^{\mu\nu}-\delta^{\mu\nu}\delta N\right)A_{\mu\nu}\ ,
\end{align}
where we have made use of the definitions \eqref{eq:SRG_redefinition of fields_N} and \eqref{eq:SRG_redefinition of fields_V}. Accordingly, we can now write down the total quadratic action by substituting  \eqref{eq:quadratic_action_SRG_diagonalised} and \eqref{eq:SRG_source_action} into \eqref{eq:functional_taylor_2} which renders
\begin{align}\label{eq:total_V_and_N}
\delta^2S^{\text{SRG}}_{\text{total}}&=2\int\text{d}^2x\Big[\delta N\bar\Box\delta N-\delta V^\nu\phantom{}_\mu\partial_\nu\partial_\alpha \delta V^{\mu\alpha}\nonumber\\
&-\frac2a\left(\delta V^{\mu\nu}-\delta^{\mu\nu}\delta N\right)A_{\mu\nu}\Big]\ .
\end{align}

Let us now consider the equations of motion for the perturbations $\delta V^{\mu\nu}$ and $\delta N$. Variation of the quadratic action \eqref{eq:total_V_and_N} with respect to $\delta N$, gives us
\begin{align}\label{eq:eom_N}
\bar\Box\delta N=-\frac1a\delta^{\mu\nu}A_{\mu\nu}\ .
\end{align}
As already mentioned, the purpose of introducing the source action $S_{\text{source}}$ is to generate the linearised spherically-reduced Schwarzschild solution given by equations \eqref{eq:SRG_linearised_Schwarzschild_local_xi} and \eqref{eq:SRG_linearised_Schwarzschild_local_dilaton} provided the smallness condition \eqref{eq:smallness_general} is satisfied. We will now show that this choice of a source action allows us to obtain the linearised spherically-reduced Schwarzschild solution. In addition, since the linearised spherically-reduced Schwarzschild solution is singular at $r=0$, we consider the case where $b=0$ in equation \eqref{eq:def_A_mu_nu}. By substituting equations \eqref{eq:SRG_redefinition of fields_N} and \eqref{eq:def_A_mu_nu} into equation \eqref{eq:eom_N} and taking $b=0$, we obtain
\begin{align}
\label{eq:SRG_eom}
\Delta\left(\bar\Phi\delta f\right)=-\frac{4M}a\,\delta'(r)\ ,
\end{align}
where the Laplacian is $\Delta:=\partial_r^2$. We now proceed to solve this equation for the perturbed metric $\delta f$ by resorting to its Fourier transform, $\mathcal{F}$. Thus,
\begin{align}
k^2\mathcal{F}\left\{\bar\Phi\delta f\right\}=\frac{4M}{\sqrt{2\pi a^2}}\int\text{d}r\ &{\text e}^{-ikr}\delta'(r)\ .
\end{align}
Integrating by parts and dividing the resulting expression by $k^2$ yields the following solution in Fourier space
\begin{align}\label{eq:SRG_fourier_transform}
\mathcal{F}\left\{\bar\Phi \delta f\right\}=\frac{4iM}{ak\sqrt{2\pi}}\ .
\end{align}
Implementing the inverse Fourier tranform on the last expressions results in
\begin{align}\label{eq:SRG_reverse_fourier}
\bar\Phi \delta f=-\frac{2M}{\pi ia}\int\text{d}k\frac{{\text e}^{ikr}}{k}\ .
\end{align}
The integral returns a value of $i\pi$ for $r>0$ while returning $-i\pi$ for $r<0$. By dividing both sides by $\bar\Phi=ar$ we obtain the solution
\begin{align}\label{eq:SRG_local_solution}
\delta f=-\frac{2M}{a^2|r|}\ .
\end{align}

It now remains to compute the perturbed dilaton field. Through the variation of equation \eqref{eq:total_V_and_N} with respect to $\delta V^{\mu\nu}$ one can obtain
\begin{align}
\partial^\alpha\partial_{(\mu}\delta V_{\nu)\alpha}=-\frac1aA_{\mu\nu}\ .
\end{align}
Substituting the definition \eqref{eq:SRG_redefinition of fields_V} into the above gives
\begin{align}
\partial_\mu\partial_\nu\delta\Phi+\delta_{\alpha(\mu}\partial_{\nu)}\partial^\alpha\delta N=-\frac{1}{a}A_{\mu\nu}\ .
\end{align}
By seeking static solutions and making use of equation \eqref{eq:eom_N}, the last expression reduces to
\begin{align}\label{eq:perturbed_dilaton_eom_SRG}
\Delta\delta\Phi=0\ ,\ \ \ \ \implies\ \ \ \delta\Phi=c_1r+c_2\ ,
\end{align}
where $c_1$ and $c_2$ are constants of integration. By invoking the constraint equation \eqref{eq:schwarzschild_type_gauge_constraint_1_v2}, we find that $c_2=0$. On the other hand, through an appropriate coordinate transformation, one can show that the parameter $c_1$ simply results in a constant rescaling of the metric. Therefore, without the loss of generality, we can safely set $c_1=0$. The perturbed dilaton is then $\delta\Phi=0$. We also note that the second constraint equation \eqref{eq:schwarzschild_type_gauge_constraint_2_v2} is satisfied since the solution is static.

Let us now examine this solution in the space-time region for which the smallness conditions are satisfied. Since $\delta\Phi=0$, equation \eqref{eq:smallness_general} is satisfied and we need only consider the perturbed metric $\delta f$. The smallness condition for the perturbed metric is $|\delta f|\ll1$. For the solution \eqref{eq:SRG_local_solution}, the smallness condition is satisfied whenever $|r|\gg2M/a^2$. Consequently for $r\gg2M/a^2$, the metric takes the form
\begin{align}
\text{d}s^2\approx-\left(1-\frac{2M}{a^2r}\right)\text{d}t^2+\left(1+\frac{2M}{a^2r}\right)\text{d}r^2\ ,
\end{align}
which is indeed the linearised spherically-reduced Schwarzschild solution.

We have thus shown that the source action \eqref{eq:SRG_source_action} can be used to generate the linearised spherically-reduced Schwarzschild solution in the context of the diagonalised theory provided the smallness conditions hold. In the following subsection, we construct a non-local modification to the SRG theory and use the same source action to determine if the solution \eqref{eq:SRG_local_solution} is modified.
\subsection{Ghost-free infinite-derivative SRG gravity}
The diagonalised quadratic action \eqref{eq:quadratic_action_SRG_diagonalised} of the local SRG theory implies that its non-local modification does not admit any additional degrees of freedom if its quadratic action is of the form
\begin{align}\label{eq:quadratic_action_SRG_non_local}
\delta^2S_{\text{non-local}}^{\text{SRG}}=4\int\text{d}^2x\big[\delta N&a(\bar\Box)\bar\Box\delta N^{\mu\nu}\nonumber\\
&-\delta V^\nu_\mu c(\bar\Box)\phantom{}\partial_\nu \partial_\alpha\delta V^{\alpha\mu}\big]\ ,
\end{align}
where $a(\bar\Box)$ and $c(\bar\Box)$ contain infinitely many derivatives and are analytic with no zeros. It is through these infinite-derivative operators that non-locality is introduced. By making use of this non-local quadratic action, we wish to study how the local solution \eqref{eq:SRG_local_solution} would be modified. To study this we consider the case where $a(\bar\Box)=c(\bar\Box)={\text e}^{-\ell^2\bar\Box}$ and $\ell\geq0$ is the \textit{length scale of non-locality}.

The non-local analogue of equation \eqref{eq:total_V_and_N} can be obtained by replacing the local quadratic action $\delta^2S_{\text{local}}$ with the non-local quadratic action $\delta^2S_{\text{non-local}}$ given in equation \eqref{eq:quadratic_action_SRG_non_local}. By making this substitution, we have
\begin{align}\label{eq:total_V_and_N_non_local}
\delta^2S_{\text{total}}^
{\text{SRG}}
&=2\int\text{d}^2x\Big[\delta Na(\bar\Box)\bar\Box\delta N-\delta V^\nu\phantom{}_\mu c(\bar\Box)\partial_\nu\partial_\alpha \delta V^{\alpha\mu}\nonumber\\
&-\frac2a\left(\delta V^{\mu\nu}-\delta^{\mu\nu}\delta N\right)A_{\mu\nu}\Big]\ .
\end{align}
Through the variation of this total quadratic action \eqref{eq:total_V_and_N_non_local} with respect to $\delta N$, we find the following non-local analogue to equation \eqref{eq:eom_N}
\begin{align}
{\text{e}}^{-\ell^2\bar\Box}\bar\Box\delta N=-\frac1a\delta^{\mu\nu}A_{\mu\nu}\ ,
\end{align}
after choosing $a(\bar\Box)={\text e}^{-\ell^2\bar\Box}$. Upon the substitution of the definitions \eqref{eq:SRG_redefinition of fields_N} and \eqref{eq:def_A_mu_nu} with $b=0$ into the last expression, we find
\begin{align}\label{eq:SRG_eom_non_local}
{\text e}^{-\ell^2\Delta}\Delta\left(\bar\Phi\delta f\right)=-\frac{4M}{a}\delta'(r)\ ,
\end{align}
which holds for static solutions. We can solve this equation of motion by first implementing the Fourier transform, dividing the resulting expression by $k^2{\text e}^{\ell^2k^2}$ and then applying the inverse Fourier transform. By doing this, we find the non-local analogue of equation \eqref{eq:SRG_reverse_fourier}
\begin{align}\label{eq:SRG_non_local_fourier}
\bar\Phi\delta f=-\frac{2M}{i\pi a}\int\text{d}k\frac{{\text e}^{ikr-\ell^2k^2}}{k}\ .
\end{align}
We can write this integral as 
\begin{align}
\bar\Phi\delta f&=-\frac{2M}{a\pi}\int^\infty_{-\infty}\text{d}k\ {\text e}^{-\ell^2k^2}\int^r_0\text{d}u\ {\text e}^{iku}\nonumber\\
&=-\frac{2M}{a\pi}\int^r_0\text{d}u\int^\infty_{-\infty}\text{d}k\ {\text e}^{-\ell^2\left(k-\frac{iu}{2\ell^2}\right)^2-\frac{u^2}{4\ell^2}}\ .
\end{align}
Evaluating it over $k$ yields
\begin{align}
\bar\Phi\delta f=-\frac{2M}{a\ell\sqrt\pi}\int^r_0\text{d}u\ {\text e}^{-\frac{u^2}{4\ell^2}}\ .
\end{align}
Performing the change of variables $v=u/2\ell$ gives us
\begin{align}
\bar\Phi\delta f=-\frac{4M}{a\sqrt\pi}\int^{r/2\ell}_0\text{d}v\ {\text e}^{-v^2}\ ,
\end{align}
where the integral above is nothing more than the error function $\frac{\sqrt\pi}2\text{Erf}(r/2\ell)$. Dividing both sides by the background dilaton field $\bar\Phi=ar$, we find the non-local modification to \eqref{eq:SRG_local_solution}
\begin{align}\label{eq:SRG_non_local_solution}
\delta f=-\frac{2M}{a^2r}\text{Erf}\left(\frac{r}{2\ell}\right)\ .
\end{align}
It is worth noting that the $1/r$ nature appearing in the linearised Schwarzschild solution of GR is also resolved by the error function in IDG \cite{Biswas:2011ar}. As remarked in the aforesaid reference, when $r\rightarrow\infty$ the error function returns a value of unity implying that the non-local solution has the same asymptotic limit as the local solution. In addition, this is also the case when we send $r\rightarrow-\infty$. On the other hand, when we send $r\rightarrow0$, the error function behaves as $\text{Erf}\left(r/2\ell\right)\sim r/2\ell$ thus resolving the singularity in the linearised regime. It is also important to note that in the limit $\ell\rightarrow0$ the error function approaches $1$ for $r>0$ and $-1$ for $r<0$; recovering the local solution as expected. For the perturbed dilaton, we find that the non-local analogue of equation \eqref{eq:perturbed_dilaton_eom_SRG} is
\begin{align}
{\text e}^{-\ell^2\Delta}\Delta\delta\Phi=0\ ,
\end{align}
which once again yields $\delta\Phi=c_1r+c_2$. We can safely set $c_1=0$ using the same argument as for the local case. In addition, by examining the constraint equation \eqref{eq:schwarzschild_type_gauge_constraint_1_v2} to first order and invoking the smallness conditions, one can show that $c_2=0$. The perturbed dilaton therefore takes on the same form as that of the local case, i.e., $\delta\Phi=0$. The second constraint equation \eqref{eq:schwarzschild_type_gauge_constraint_2_v2} is satisfied since the solution is static.

\begin{figure}
  \includegraphics[width=1\columnwidth]{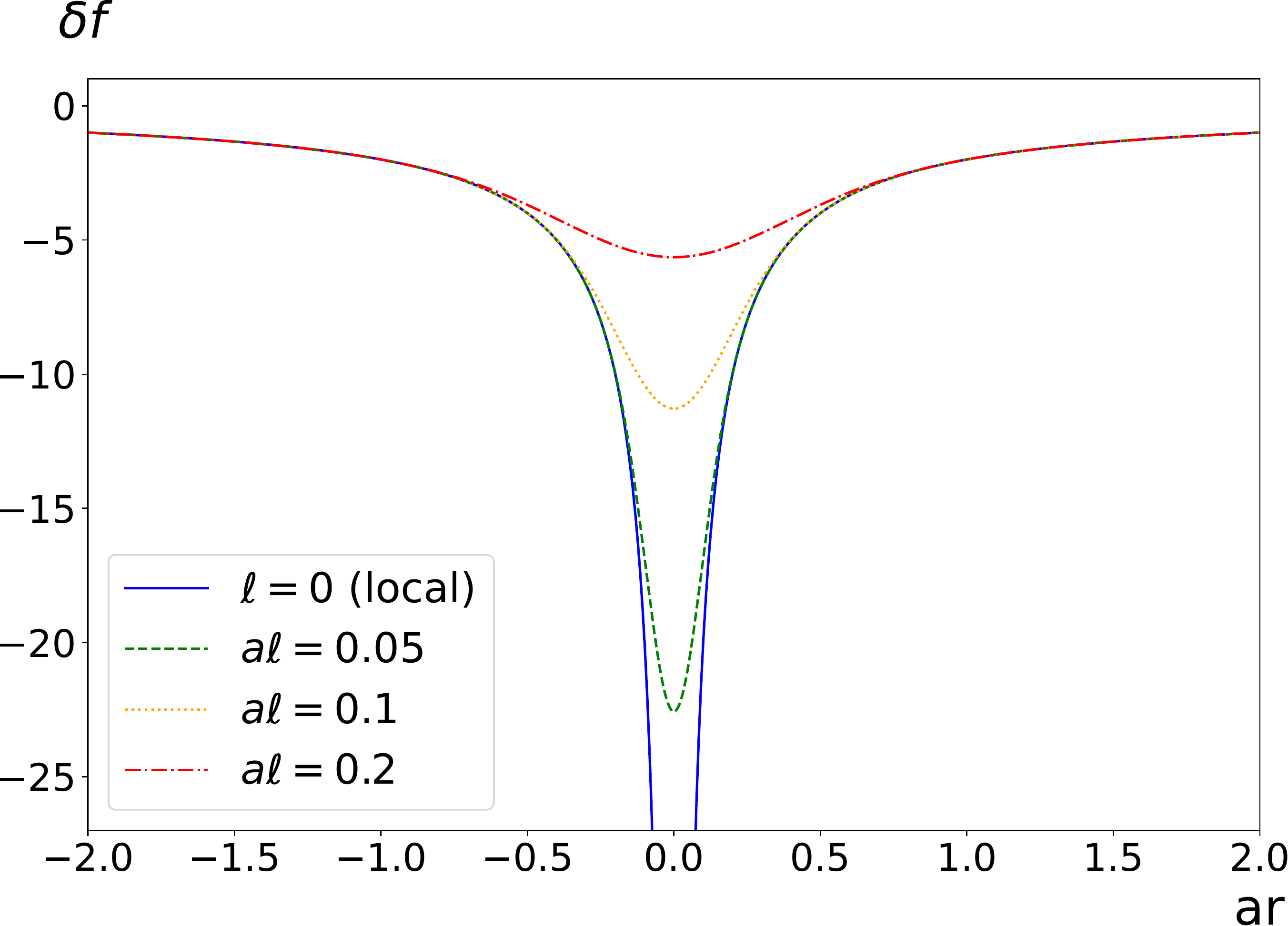}
\caption{Perturbed metric $\delta f$ radial dependence for the SRG theory as given by equations \eqref{eq:SRG_local_solution} and \eqref{eq:SRG_non_local_solution} for the local and non-local cases respectively. The solid blue curve shows the local case, i.e., $\ell=0$. The remaining three curves correspond to non-local solutions with the dashed green, dotted orange and dash-dotted red curves corresponding to $a\ell=0.05$, $a\ell=0.1$ and $a\ell=0.2$ respectively. In producing these plots, we have set $a=M=1$.}
\label{fig:delta_f}
\end{figure}

\begin{figure}
  \includegraphics[width=1\columnwidth]{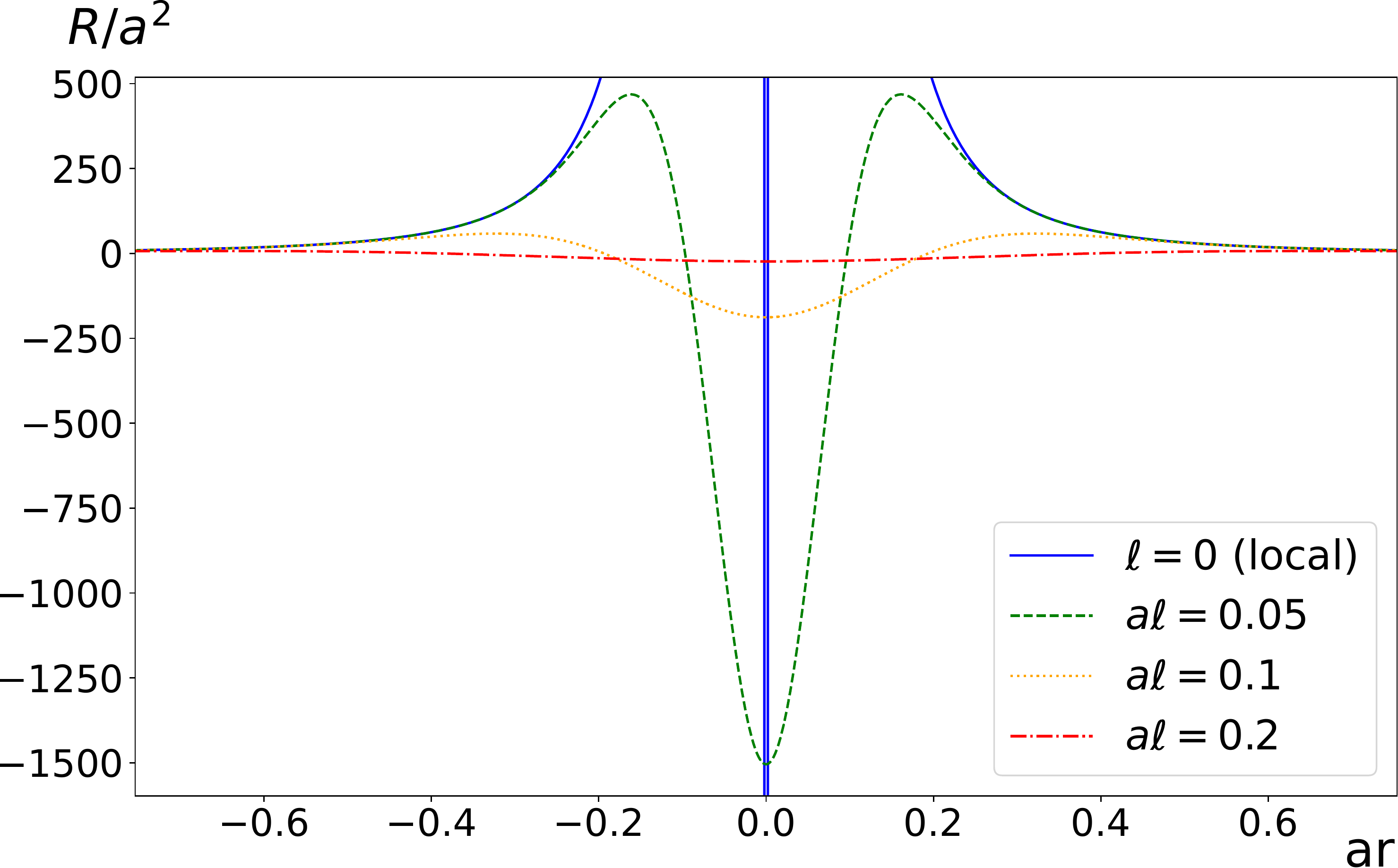}
\caption{Radial dependence of the Ricci scalar $R/a^2$ for the SRG theory when calculated to first order in $\delta f$ using $R=-\Delta\delta f+\mathcal{O}\left(\delta f^2\right)$. The solid blue curve shows the local case, i.e., $\ell=0$, which is singular at the origin. The remaining three curves correspond to non-local solutions with the dashed green, dotted orange and dash-dotted red curves corresponding to $a\ell=0.05$, $a\ell=0.1$ and $a\ell=0.2$ respectively. The vertical line through the origin is included to show the distributional character of the Ricci scalar for the local case as well as how it is regularised in the non-local cases. In producing these plots, we have set $a=M=1$.}
\label{fig:R_srg}
\end{figure}

In Figures \ref{fig:delta_f} and \ref{fig:R_srg} we plot the perturbed metric $\delta f$ and Ricci scalar $R$ respectively for the local case as well as three non-local scenarios. The solid blue curves correspond to the local ($\ell=0$) case whereas the dashed green, dotted orange and dash-dotted red curves correspond to the non-local cases with length scale of non-locality parameters $a\ell=0.05$, $a\ell=0.1$ and $a\ell=0.2$ respectively. Given the perturbed metric $\delta f$ we compute the Ricci scalar to first order, i.e., by using the expression $R=-\Delta\delta f+\mathcal{O}\left(\delta f^2\right)$.

\section{CGHS gravity}
\label{sec:IDG_CGHS}

We now wish to follow a similar procedure to study ghost-free infinite-derivative modifications of the CGHS theory.
\subsection{Conformal gauge}
The vacuum CGHS theory is given by the action \eqref{eq:action_with_Phi} with $k=1$, $a=0$ and $\lambda$ left unspecified \cite{Callan:1992rs}. While for the case of SRG we worked in the Schwarzschild-type gauge, for the CGHS theory we shall implement the conformal gauge, i.e., we fix
\begin{align}\label{eq:conformal_gauge_metric}
g_{\mu\nu}={\text e}^{2w}\eta_{\mu\nu}\ ,
\end{align}
where $\eta_{\mu\nu}$ is the Minkowski metric in Cartesian coordinates and we refer to $w$ as the \textit{conformal scalar}.

In the conformal gauge and starting from equation \eqref{eq:action_with_Phi}, we write the CGHS action as
\begin{align}\label{eq:CGHS_conformal_gauge}
S^{\text{CGHS}}_{\text{local}}=4\int\textup{d}^2x\left[\lambda^2\Phi^2{\text e}^{2w}+\left(\partial\Phi\right)^2-\frac12\Phi^2\Box w\right]\ ,
\end{align}
where we have used the fact that, in this scenario, the Ricci scalar becomes
\begin{align}
\label{eq:Ricci_scalar_conformal_gauge}
R=-2{\text e}^{-2w}\Box w\ ,
\end{align}
where $\Box:=\eta^{\mu\nu}\partial_\mu\partial_\nu$ is the d'Alembertian operator in Minkowski space-time $\mathbb{R}^{1,1}$. We emphasise that we now raise and lower indices using the Minkowski metric $\eta^{\mu\nu}$. In addition, we have the conformal gauge constraint equations
\begin{align}\label{eq:conformal_gauge_constraint_equation_CGHS_v2}
\delta^{\mu\nu}\left(4\partial_\mu\Phi\partial_\nu\Phi+2\partial_{\mu}w\partial_{\nu}\Phi^2-\partial_\mu\partial_\nu\Phi^2\right)=0\ ,
\end{align}
and
\begin{align}\label{eq:conformal_gauge_constraint_equation_2_CGHS_v2}
4\partial_\mu\Phi\partial_\nu\Phi+2\partial_{(\mu}w\partial_{\nu)}\Phi^2-\partial_\mu\partial_\nu\Phi^2=0\ ,\ \ \ \ (\mu\neq\nu)\ .
\end{align}
Variation of the gauge-fixed action \eqref{eq:CGHS_conformal_gauge} yields
\begin{align}\label{eq:variation_CGHS_conformal_gauge}
\delta S^{\text{CGHS}}_{\text{local}}=4&\int\text{d}^2x\bigg(2\lambda^2\Phi\delta\Phi {\text e}^{2w}+2\lambda^2\Phi^2{\text e}^{2w}\delta w\nonumber\\
&+2\partial_\nu\Phi\partial^\nu\delta\Phi-\Phi\delta\Phi\Box w-\frac12\Phi^2\Box \delta w\bigg)\ .
\end{align}
As mentioned in Section \ref{sec:degrees_of_freedom}, in addition to the two redundancies appearing as a result of diffeomorphism invariance, there is the additional symmetry described by equation \eqref{eq:CGH_additional_symmetry}. Let us verify that the CGHS action is indeed left invariant under this transformation. In the conformal gauge, equation \eqref{eq:CGH_additional_symmetry} reads
\begin{align}\label{eq:CGHS_symmetry_conformal_gauge}
\delta w=-\frac{\varepsilon}{\Phi^2}\ ,\ \ \ \ \ \ \ \ \ \ \delta\Phi=\frac{\varepsilon}{\Phi}\ .
\end{align}
Substituting \eqref{eq:CGHS_symmetry_conformal_gauge} into equation \eqref{eq:variation_CGHS_conformal_gauge} gives us
\begin{align}\label{eq:CGHS_vary_sub_in_symmetry}
\delta S^{\text{CGHS}}_{\text{local}}=4\varepsilon\int\text{d}^2x\bigg[2\partial^\nu&\Phi\partial_\nu\left(\frac{1}{\Phi}\right)\nonumber\\
&+\frac12\Phi^2\Box\left(\frac1{\Phi^2}\right)-\Box w\bigg]\ .
\end{align}
The last step required to verify that equation \eqref{eq:CGHS_symmetry_conformal_gauge} is a symmetry of the CGHS action is to show that the integrand in the above is a total derivative. To accomplish this, we note that
\begin{align}
\partial^\nu\left(\Phi^2\partial_\nu\frac{1}{\Phi^2}\right)=\Phi^2\Box\left(\frac{1}{\Phi^2}\right)+4\partial^\nu\Phi\partial_\nu\left(\frac{1}{\Phi}\right)\ .
\end{align}
It now follows that equation \eqref{eq:CGHS_vary_sub_in_symmetry} reads
\begin{align}
\delta S^{\text{CGHS}}_{\text{local}}=-4\varepsilon\int\text{d}^2x\ \partial^\nu\left(\frac{\partial_\nu\Phi}{\Phi}+\partial_\nu w\right)\ ,
\end{align}
showing that the integrand is a total derivative and therefore the CGHS action is invariant under the transformation \eqref{eq:CGHS_symmetry_conformal_gauge}.
\subsection{Diagonalisation of the quadratic action}
We now turn our attention to examining the quadratic CGHS action in the conformal gauge. The functional derivative of the action \eqref{eq:CGHS_conformal_gauge} with respect to the dilaton field yields
\begin{align}\label{eq:CGHS_eom_dilaton}
\frac{\delta S^{\text{CGHS}}_{\text{local}}}{\delta\Phi}=8\left(\lambda^2\Phi {\text e}^{2w}-\Box\Phi-\Phi\Box w/2\right)\ ,
\end{align}
while the functional derivative with respect to the conformal scalar leads to
\begin{align}\label{eq:CGHS_eom_metric}
\frac{\delta S^{\text{CGHS}}_{\text{local}}}{\delta w}=8\left[\lambda^2\Phi^2 {\text e}^{2w}-\left(\partial\Phi\right)^2/2-\Phi\Box\Phi/2\right]\ .
\end{align}
The corresponding field equations are obtained by setting both equations \eqref{eq:CGHS_eom_dilaton} and \eqref{eq:CGHS_eom_metric} equal to zero. To study the quadratic action for this theory, we perturb the dilaton field and conformal scalar around some background solution $(\bar\Phi,\bar w)$. That is, we write
\begin{align}\label{eq:CGHS_perturb_dilaton_metric}
w=\bar w+\delta w\ ,\ \ \ \ \ \ \ \ \ \
\Phi=\bar\Phi+\delta\Phi\ ,
\end{align}
where $(\bar\Phi,\bar w)$ solves the equations of motion and $\delta\Phi$ and $\delta w$ are the perturbations. In order for the quadratic action to be a valid approximation, we require that the smallness conditions are satisfied. The smallness condition for the dilaton field remains $|\delta\Phi|\ll|\bar\Phi|$. To first order in $\delta w$ the metric tensor is
\begin{align}
g_{\mu\nu}={\text e}^{2\bar w}\eta_{\mu\nu}\left[1+2\delta w+\mathcal{O}\left(\delta w^2\right)\right]\ ,
\end{align}
which follows from equation \eqref{eq:conformal_gauge_metric}. From this expression, we can extract the smallness condition for the perturbed conformal scalar
\begin{align}\label{eq:CGHS_perturb_dilaton_metric_smallness}
|\delta w|\ll\frac12\ .
\end{align}
Provided that these two conditions are satisfied, the quadratic action will be a valid approximation. In equation \eqref{eq:quadratic_action_definition} we defined the quadratic action for a general background metric $\bar g_{\mu\nu}$. In the conformal gauge, equation \eqref{eq:quadratic_action_definition} becomes
\begin{align}\label{eq:quadratic_action_definition_CGHS}
\delta^2S^{\text{CGHS}}_{\text{local}}&:=\int\text{d}^2x\ \text{d}^2x'\Bigg[\delta\Phi(x)\delta\Phi(x')\frac{\delta^2S_{\text{local}}}{\delta\Phi(x)\delta\Phi(x')}\Bigg|_{\left(\bar w,\bar\Phi\right)}\nonumber \\
&+\delta\Phi(x)\delta w(x')\frac{\delta^2S_{\text{local}}}{\delta\Phi(x)\delta w(x')}\Bigg|_{\left(\bar w,\bar\Phi\right)}\nonumber\\
&+\delta w(x)\delta\Phi(x')\frac{\delta^2S_{\text{local}}}{\delta w(x)\delta\Phi(x')}\Bigg|_{\left(\bar w,\bar\Phi\right)}\nonumber\\
&+\delta w(x)\delta w(x')\frac{\delta^2S_{\text{local}}}{\delta w(x)\delta w(x')}\Bigg|_{\left(\bar w,\bar\Phi\right)}\Bigg]\ .
\end{align}
Let us now compute the quadratic action for the CGHS theory in the conformal gauge for a general background solution. The second-order functional derivatives being
\begin{align}
\frac{\delta^2S^{\text{CGHS}}_{\text{local}}}{\delta\Phi(x')\delta\Phi(x)} &=8\left(\lambda^2{\text e}^{2w}-\Box-\Box w/2\right)
\delta^{(2)}(x-x')\ ,\label{eq:CGHS_Phi_Phi}
\\
\frac{\delta^2S^{\text{CGHS}}_{\text{local}}}{\delta w(x')\delta\Phi(x)} &=8\left(2\lambda^2 {\text e}^{2w}\Phi-\frac{1}{2}\Phi\Box\right)\delta^{(2)}(x-x')\ ,\label{eq:CGHS_Phi_w}
\\
\frac{\delta^2S^{\text{CGHS}}_{\text{local}}}{\delta\Phi(x')\delta w(x)} &=8\left(2\lambda^2\Phi {\text e}^{2w}-\partial^\mu\Phi\partial_\mu-\Box\Phi/2-\frac12\Phi\Box\right)\nonumber\\
&\times\delta^{(2)}(x-x')\ ,\label{eq:CGHS_w_Phi}
\end{align}
and
\begin{align}
\frac{\delta^2S^{\text{CGHS}}_{\text{local}}}{\delta w(x')\delta w(x)} &=16\lambda^2\Phi^2{\text e}^{2w}\delta^{(2)}(x-x')\ .\label{eq:CGHS_w_w}
\end{align}
The substitution of 
the above second-order functional derivatives when evaluated at the background solution $\left(\bar w,\bar\Phi\right)$ renders equation \eqref{eq:quadratic_action_definition_CGHS} into
\begin{align}\label{eq:CGHS_quadratic_action}
&\delta^2S^{\text{CGHS}}_{\text{local}}=8\int\textup{d}^2x\ \bigg\{\delta\Phi\left[\lambda^2{\text e}^{2\bar w}-\Box-\Box\bar w/2\right]\delta\Phi\nonumber\\
&+\delta w\left[2\lambda^2{\text e}^{2\bar w}\bar\Phi-\partial^\mu\bar\Phi\partial_\mu-\Box\bar\Phi/2-\bar\Phi\Box/2\right]\delta\Phi\nonumber\\
&+\delta\Phi\left[2\lambda^2\bar\Phi {\text e}^{2\bar w}-\bar\Phi\Box/2\right]\delta w+2\lambda^2\delta w{\bar\Phi}^2{\text e}^{2\bar w}\delta w\bigg\}\ ,
\end{align}
which can be further simplified
by using the equation of motion \eqref{eq:CGHS_eom_dilaton} in the first bracket and integrating by parts in the second bracket. Doing this brings the quadratic action to a more tractable form
\begin{align}\label{eq:CGHS_quadratic_action_simplified}
&\delta^2S^{\text{CGHS}}_{\text{local}}=8\int\textup{d}^2x\ \bigg[2\lambda^2\delta w{\bar\Phi}^2{\text e}^{2\bar w}\delta w\nonumber\\
&+\delta\Phi\left(\frac{\Box\bar\Phi}{\bar\Phi}-\Box\right)\delta\Phi+\delta\Phi\left(4\lambda^2\bar\Phi {\text e}^{2\bar w}-\bar\Phi\Box\right)\delta w\bigg]\ .
\end{align}
From here, we proceed with the diagonalisation. First, we define the field
\begin{align}\label{eq:CGHS_define_psi}
\delta\psi:=\delta w+\frac{\delta \Phi}{\bar\Phi}\ ,
\end{align}
which is a linear combination of the perturbed metric and dilaton. Substituting this definition for the perturbed conformal scalar $\delta w$ in the quadratic action \eqref{eq:CGHS_quadratic_action_simplified} results in
\begin{align}\label{eq:CGHS_quadratic_action_with_psi}
&\delta^2S^{\text{CGHS}}_{\text{local}}=8\int\textup{d}^2x\Bigg\{2\lambda^2\bar\Phi^2{\text e}^{2\bar w}\left[\frac{\delta\Phi^2}{\bar\Phi^2}+\delta\psi^2\right]
-\delta\Phi\bar\Phi\Box\delta\psi\nonumber\\
&+\delta\Phi\bigg[-4\lambda^2\delta\Phi {\text e}^{2\bar w}+\bar\Phi\Box\left(\frac{\delta\Phi}{\bar\Phi}\right)+\left(\frac{\Box\bar\Phi}{\bar\Phi}-\Box\right)\delta\Phi\bigg]\Bigg\}\ .
\end{align}
At this stage we can make use of the fact that the following combination is a total derivative
\begin{align}\label{eq:Phi_identity}
\Phi\Box\Phi+\frac{\Phi^2\left(\partial\bar\Phi\right)^2}{\bar\Phi\phantom{}^2}-\Phi\bar\Phi\Box\left(\frac{\Phi}{\bar\Phi}\right)=\partial_\nu\left[\frac{\Phi^2\partial^\nu\bar\Phi}{\bar\Phi}\right]\ ,
\end{align}
so equation \eqref{eq:CGHS_quadratic_action_with_psi} can be rewritten as
\begin{align}\label{eq:CGHS_quadratic_action_with_psi_simplified}
&\delta^2S^{\text{CGHS}}_{\text{local}}=8\int\textup{d}^2x\Bigg[-2\lambda^2\delta\Phi^2\ {\text e}^{2\bar w}+2\lambda^2\bar\Phi^2\ {\text e}^{2\bar w}\delta\psi^2\nonumber\\
&+\frac{\delta\Phi^2}{\bar\Phi^2}\partial_\nu\left(\bar\Phi\partial^\nu\bar\Phi\right)-\delta\Phi\bar\Phi\Box\delta\psi-\partial_\nu\left(\frac{\delta\Phi^2\partial^\nu\bar\Phi}{\bar\Phi}\right)\Bigg]\ .
\end{align}
The last term in the integrand is a total derivative and therefore has no contribution. Now, since the background solution $\left(\bar w,\bar\Phi\right)$ satisfies the equations of motion, we can make use of equation \eqref{eq:CGHS_eom_metric} in the above expression in order to obtain the following, much simpler, form of the quadratic CGHS action
\begin{align}\label{eq:CGHS_quadratic_action_with_psi_simplified2}
\delta^2S^{\text{CGHS}}_{\text{local}}=8\int\textup{d}^2x\big\{2\lambda^2\bar\Phi^2{\text e}^{2\bar w}\delta\psi^2-\delta\psi\Box\left(\bar\Phi\delta\Phi\right)\big\}\ .
\end{align}
We can further simplify the action if we vary equation \eqref{eq:CGHS_quadratic_action_with_psi_simplified2} with respect to the perturbation $\delta\psi$ which gives
\begin{align}\label{eq:constraint_equation_delta_w}
\delta w=-\frac{\delta\Phi}{\bar\Phi}+\frac{\Box\left(\bar\Phi\delta\Phi\right)}{4\lambda^2\bar\Phi^2 {\text e}^{2\bar w}}\ ,
\end{align}
after substituting in the definition \eqref{eq:CGHS_define_psi} for the original fields. It follows from this last expression that $\delta w$ is an auxiliary field. By substituting this constraint equation into the quadratic action \eqref{eq:CGHS_quadratic_action_with_psi_simplified2} and defining
\begin{align}\label{eq:CGHS_define_epsilon}
\delta\chi:=\frac{\Box\left(\bar\Phi\delta\Phi\right)}{\sqrt8\lambda\bar\Phi {\text e}^{\bar w}}\ ,
\end{align}
we can write schematically the  diagonalised quadratic CGHS action as
\begin{align}\label{eq:CGHS_quadratic_action_final}
\delta^2S^{\text{CGHS}}_{\text{local}}=-8\int\textup{d}^2x\ \delta\chi^2\ ,
\end{align}
which describes the one propagating off-shell degree of freedom. Equation \eqref{eq:CGHS_quadratic_action_final} is the desired diagonalised quadratic action in terms of the redefined field $\delta\chi$.
%
In Section \ref{sec:CGHS_idg} we shall return to the latter expression in order to construct a ghost-free infinite derivative modification of the CGHS quadratic action. Before doing this, let us briefly discuss the CGHS BH solution in the conformal gauge.

\subsection{CGHS BH solution} \label{subsec:BHS}
Let us now study the CGHS BH solution \cite{Callan:1992rs} in the conformal gauge. We will first discuss the full local CGHS BH solution in the conformal gauge and then move on to discuss the solution in the linearised regime.
\subsubsection{General solution}
By dividing the equation of motion \eqref{eq:CGHS_eom_metric} by $\Phi$ and then subtracting this from equation \eqref{eq:CGHS_eom_dilaton}, one obtains
\begin{align}\label{eq:CGHS_on_shell_gauge_1}
\Box\left(w+\ln\Phi\right)=0\ .
\end{align}
At this point, we can remove the redundancy arising from \eqref{eq:CGH_additional_symmetry} by fixing on-shell\footnote{Let us note that such a fixing can only be done on-shell, because if performed off-shell, information about the field dynamics would be lost.} the following
\begin{align}\label{eq:CGHS_metric_to_dilaton}
{\text e}^w\Phi={\text e}^{\lambda r}\ ,
\end{align}
which allows for equation \eqref{eq:CGHS_on_shell_gauge_1} to be satisfied. With this choice of gauge, the equation of motion \eqref{eq:CGHS_eom_metric} gives us
\begin{align}\label{eq:CGHS_derive_BH_dilaton_de}
\Box\Phi^2=4\lambda^2{\text e}^{2\lambda r}\ .
\end{align}
Equation \eqref{eq:CGHS_derive_BH_dilaton_de} is easily solvable to find an expression for the dilaton
\begin{align}\label{eq:CGHS_deriv_BH_dilaton_solution}
\Phi^2={\text e}^{2\lambda r}+E\ ,
\end{align}
where $E\in\mathbb{R}$ is a constant. From equation \eqref{eq:CGHS_on_shell_gauge_1}, it now follows that the metric is of the form
\begin{align}\label{eq:CGHS_metric_C}
\text{d}s^2=\frac{-\text{d}t^2+\text{d}r^2}{1+E{\text e}^{-2\lambda r}}\ ,
\end{align}\label{eq:CGHS_conformal_scalar_C}
with the conformal scalar written as
\begin{align}
w=-\frac12\ln\left(1+E{\text e}^{-2\lambda r}\right)\ .
\end{align}
As it is done in \cite{Callan:1992rs,Grumiller:2002nm}, one can show that the constant $E$ is related to the ADM mass $M$, which has dimensions of length$\phantom{}^{-1}$, through $|E|=M/\lambda$. When $E$ is positive, the metric \eqref{eq:CGHS_metric_C} describes the region exterior to the CGHS BH with $(t,r)\in\left(-\infty,\infty\right)\times\left(-\infty,\infty\right)$. When $E=-M/\lambda$ the metric describes the interior region of the BH with coordinate range $(t,r)\in\left(-\infty,\infty\right)\times\left(-\infty,\frac{1}{2\lambda}\ln\left(M/\lambda\right)\right)$ where $r=\frac{1}{2\lambda}\ln\left(M/\lambda\right)$ is the singularity and $r=-\infty$ is the horizon.
\subsubsection{Linearised local solution}
Setting $E=M/\lambda$ in equation \eqref{eq:CGHS_metric_C} gives the metric that describes the exterior region of the CGHS BH with mass $M$ for $r\in\left(-\infty,\infty\right)$ and horizon located at $r=-\infty$. To first order in $M/\lambda$, the conformal scalar is
\begin{align}\label{eq:CGHS_conformal_scalar_linearised}
w=-\frac{M}{2\lambda}{\text e}^{-2\lambda r}+\mathcal{O}\left(\frac{M^2}{\lambda^2}\right)\ ,
\end{align}
while the dilaton field reads
\begin{align}\label{eq:CGHS_dilaton_linearised}
\Phi={\text e}^{\lambda r}+\frac{M}{2\lambda}{\text e}^{-\lambda r}+\mathcal{O}\left(\frac{M^2}{\lambda^2}\right)\ .
\end{align}
From this expansion, the background fields correspond to the linear dilaton solution
\begin{align}\label{eq:CGHS_background}
\bar w=0\ , \ \ \ \ \text{and}\ \ \ \ \bar\Phi={\text e}^{\lambda r}\ ,
\end{align}
while the perturbed fields are
\begin{align}\label{eq:CGHS_local_perturbed_fields}
\delta w=\frac{M}{2\lambda}{\text e}^{-2\lambda r}\ , \ \ \ \ \text{and}\ \ \ \ \delta\Phi=-\frac{M}{2\lambda}{\text e}^{-\lambda r}\ .
\end{align}
It is clear that the smallness conditions are satisfied provided $r\gg\frac{1}{2\lambda}\ln\frac{M}{\lambda}$.

\subsection{Local diagonalised theory with source action}
Let us consider the source action \eqref{eq:general_source_action_sp} for the case of the local CGHS theory in the conformal gauge. By setting $a=0$ and specifying the conformal gauge, the source action reads
\begin{align}\label{eq:source_action_CGHS_for_conclusion}
S^{\text{CGHS}}_{\text{source}}=\frac1{2\lambda}\int\text{d}^2x\frac{{\text e}^{2w}\Phi^2{\bar g}^{\mu\nu}A_{\mu\nu}}{\bar\Phi^2}\ ,
\end{align}
whose variation renders
\begin{align}\label{eq:source_action_CGHS_conformal_gauge_not_evaluated}
\delta S^{\text{CGHS}}_{\text{source}}=\frac1{\lambda}\int\text{d}^2x\frac{{\text e}^{2w}\Phi\left(\Phi\delta w+\delta\Phi\right){\bar g}^{\mu\nu}A_{\mu\nu}}{\bar\Phi^2}\ .
\end{align}
At this stage we wish to verify that this source action is invariant under the transformation \eqref{eq:CGHS_symmetry_conformal_gauge}. From equation \eqref{eq:CGHS_symmetry_conformal_gauge} it is readily seen that $\Phi\delta w+\delta\Phi=0$ which implies the vanishing of equation \eqref{eq:source_action_CGHS_conformal_gauge_not_evaluated}. Consequently, the symmetry \eqref{eq:CGH_additional_symmetry} is preserved when introducing the source action \eqref{eq:source_action_CGHS_for_conclusion}. Moreover, when evaluating the above expression at the background fields $\Phi=\bar\Phi={\text e}^{\lambda r}$ and $w=\bar w=0$ we have
\begin{align}\label{eq:source_action_CGHS_conformal_gauge}
\delta S^{\text{CGHS}}_{\text{source}}=\frac1{\lambda}\int\text{d}^2x\left(\delta w+\frac{\delta\Phi}{\bar\Phi}\right)\eta^{\mu\nu}A_{\mu\nu}\ ,
\end{align}
which in terms of the field $\delta\chi$ as defined in  \eqref{eq:CGHS_define_epsilon}, renders the last expression into
\begin{align}\label{eq:CGHS_source_action}
\delta S^{\text{CGHS}}_{\text{source}}=\frac{1}{\sqrt2\lambda^2}\int\text{d}^2x\ \frac{\delta\chi\eta^{\mu\nu}A_{\mu\nu}}{\bar\Phi}\ .
\end{align}
Since the total quadratic action to be considered in the following would be of the form given in \eqref{eq:functional_taylor_2}, let us use equation \eqref{eq:CGHS_quadratic_action_final} for $\delta^2S_{\text{local}}$ as well as equation \eqref{eq:CGHS_source_action} for $\delta S_{\text{source}}$. Thus, the total quadratic action becomes
\begin{align}
\label{eq:CGHS_total_action_quadratic_plus_source}
\delta^2S^{\text{CGHS}}_{\text{total}}=-4\int\text{d}^2x\left(\delta\chi^2-\frac{\delta\chi\eta^{\mu\nu}A_{\mu\nu}}{4\sqrt2\lambda^2\bar\Phi}\right)\ .
\end{align} 
Then, the variation of equation \eqref{eq:CGHS_total_action_quadratic_plus_source} with respect to the perturbation $\delta\chi$ yields the following equation of motion
\begin{align}\label{eq:CGHS_eom_epsilon}
\delta\chi=\frac{\eta^{\mu\nu}A_{\mu\nu}}{4\sqrt8\lambda^2\bar\Phi}\ ,
\end{align}
with the tensor field $A_{\mu\nu}$ given in equation \eqref{eq:def_A_mu_nu}. We will now show how this tensor field correctly generates the linearised CGHS BH solution provided the smallness condition is satisfied. Indeed, the substitution of equations \eqref{eq:CGHS_define_epsilon} and \eqref{eq:def_A_mu_nu} transforms \eqref{eq:CGHS_eom_epsilon} into
\begin{align}\label{eq:CGHS_lgtpo_1}
{\text e}^{-\lambda r}\Delta\left({\text e}^{\lambda r}\delta\Phi\right)=\frac{M{\text e}^{-\lambda r}}{2\lambda}\delta'(r-b)\ ,
\end{align}
provided we focus on static solutions. We now proceed to solve equation \eqref{eq:CGHS_lgtpo_1} by performing the Fourier transform, which results in
\begin{align}
\frac{1}{\sqrt{2\pi}}\int&\text{d}r\ {\text e}^{-r(ik+\lambda)}\Delta\left({\text e}^{\lambda r}\delta\Phi\right)\nonumber\\
&=\frac{M}{2\lambda\sqrt{2\pi}}\int\text{d}r\ {\text e}^{-r(ik+\lambda)}\delta'(r-b)\ .
\end{align}
Integration by parts twice on the left-hand-side and once on the right-hand-side
provides the following solution in Fourier space
\begin{align}\label{eq:CGHS_Fourier_space_solution}
\mathcal{F}\left\{\delta\Phi\right\}=\frac{M{\text e}^{-b\lambda}}{2\lambda i\sqrt{2\pi}}\frac{{\text e}^{-ikb}}{k-i\lambda}\ ,
\end{align}
whose inverse Fourier transform yields
\begin{align}
\delta\Phi=\frac{M{\text e}^{-b\lambda}}{4\pi i\lambda}\int\text{d}k\frac{{\text e}^{ik(r-b)}}{k-i\lambda}\ .
\end{align}
Upon the evaluation of the integral, we obtain the following expression for the perturbed dilaton
\begin{align}\label{eq:CGHS_solved_dilaton}
\delta\Phi=\frac{M{\text e}^{-\lambda r}}{2\lambda}\Theta(r-b)\ ,
\end{align}
where $\Theta(r-b)$ is the usual 
Heaviside step function.
We now turn our attention to obtaining an expression for the perturbed conformal scalar $\delta w$. This can be achieved by substituting equation \eqref{eq:CGHS_solved_dilaton} into the constraint equation \eqref{eq:constraint_equation_delta_w} in order to find
\begin{align}\label{eq:perturbed_metric_obtained_local}
\delta w=-\frac{M{\text e}^{-2\lambda r}}{2\lambda}\Theta(r-b)+\frac{M{\text e}^{-2\lambda r}}{8\lambda^3}\delta'(r-b)\ .
\end{align}
From equation \eqref{eq:perturbed_metric_obtained_local} we see that the perturbed metric is singular at $r=b$. In addition, for $b<\frac1{2\lambda}\ln\frac{M}{\lambda}\ll r$, equations \eqref{eq:CGHS_solved_dilaton} and \eqref{eq:perturbed_metric_obtained_local} coincide with the linearised CGHS BH solution \eqref{eq:CGHS_local_perturbed_fields}. This shows that the solution to the quadratic action \eqref{eq:CGHS_total_action_quadratic_plus_source} coincides with the linearised CGHS BH solution provided we consider the space-time region for which the smallness conditions are satisfied. It is straightforward to verify that the conformal gauge constraints \eqref{eq:conformal_gauge_constraint_equation_CGHS_v2} and \eqref{eq:conformal_gauge_constraint_equation_2_CGHS_v2} at first order are satisfied when the smallness conditions are taken into account.

\subsection{Ghost-free infinite-derivative CGHS gravity}\label{sec:CGHS_idg}

We now turn our attention to constructing infinite-derivative modifications of the CGHS theory that are ghost-free at the quadratic level. To this end, the local quadratic action \eqref{eq:CGHS_quadratic_action_final} implies that the non-local quadratic action
\begin{align}\label{eq:CGHS_quadratic_action_non_local}
\delta^2S_{\text{non-local}}^{\text {CGHS}}=-8\int\textup{d}^2x\ \delta\chi a(\Box)\delta\chi\ ,
\end{align}
admits no additional degrees of freedom provided that $a(\Box)$, which contains infinitely many derivatives, is analytic with no zeros. Thus, this will not add any new degrees of freedom off-shell, and hence no ghosts. Replacing the local quadratic action $\delta^2S_{\text{local}}$ in equation \eqref{eq:functional_taylor_2} with the non-local quadratic action $\delta^2S_{\text{non-local}}$ given above, and using the same source action as before, we find the non-local analogue of equation \eqref{eq:CGHS_total_action_quadratic_plus_source}
\begin{align}\label{eq:CGHS_total_action_quadratic_plus_source_non_local}
\delta^2S_{\text{total}}^{\text {CGHS}}=-4\int\text{d}^2x\left(\delta\chi a(\Box)\delta\chi-\frac{\delta\chi\eta^{\mu\nu}A_{\mu\nu}}{4\sqrt2\lambda^2\bar\Phi}\right)\ .
\end{align} 
Through the variation of equation \eqref{eq:CGHS_total_action_quadratic_plus_source_non_local} with respect to the perturbation $\delta\chi$ we obtain
\begin{align}
a(\Box)\delta\chi=\frac{\eta^{\mu\nu}A_{\mu\nu}}{4\sqrt8\lambda^2\bar\Phi}\ .
\end{align}
By substituting equations \eqref{eq:CGHS_define_epsilon} and \eqref{eq:def_A_mu_nu} into the above expression and seeking static solutions we obtain the non-local analogue of equation \eqref{eq:CGHS_lgtpo_1}
\begin{align}\label{eq:CGHS_lgtpo_1_non_local}
a(\Delta)\left[{\text e}^{-\lambda r}\Delta\left({\text e}^{\lambda r}\delta\Phi\right)\right]=-\frac{M{\text e}^{-\lambda r}}{2\lambda}\delta'(r-b)\ .
\end{align}
Performing the Fourier transform gives us the non-local solution in Fourier space
\begin{align}\label{eq:CGHS_Fourier_space_solution_non_local}
\mathcal{F}\left\{\delta\Phi\right\}=\frac{M{\text e}^{-b\lambda}}{2\lambda i\sqrt{2\pi}}\frac{{\text e}^{-ikb}}{a(-k^2)\left(k-i\lambda\right)}\ .
\end{align}
In order to obtain the solution in position space, we first need to specify the form of the operator $a(\Box)$. Here, we use the same choice that was used in the case of SRG and consider $a(\Box)={\text e}^{-\ell^2\Box}$ where $\ell$ is the length scale of non-locality as before. With this choice for the operator $a(\Box)$, the inverse Fourier transform of equation \eqref{eq:CGHS_Fourier_space_solution_non_local} gives
\begin{align}\label{eq:CGHS_inverse_Fourier_space_solution_non_local}
\delta\Phi=\frac{M{\text e}^{-b\lambda}}{4\pi i\lambda}\int\text{d}k\ {\text e}^{-\ell^2k^2}\frac{{\text e}^{ik(r-b)}}{k-i\lambda}\ ,
\end{align}
which corresponds to the non-local analogue of expression \eqref{eq:CGHS_Fourier_space_solution}. By evaluating the integral on the right-hand-side, we obtain the non-local solution for the perturbed dilaton field
\begin{align}\label{eq:CGHS_non_local_dilaton_solution}
\delta\Phi=\frac{M{\text e}^{\ell^2\lambda^2-\lambda r}}{4\lambda}\text{Erfc}\left(\lambda\ell+\frac{b-r}{2\ell}\right)\ ,
\end{align}
where $\text{Erfc}$ denotes the complementary error function. The interested reader seeking a derivation of equation \eqref{eq:CGHS_non_local_dilaton_solution} is directed to Appendix \ref{section:derive_CGHS_non_local_dilaton}.

Let us now consider the local limit $\ell\rightarrow0$ and verify that we recover the solution \eqref{eq:CGHS_solved_dilaton}. In this limit, we have $\lim_{\ell\rightarrow0}\ {\text e}^{\ell^2\lambda^2}=1$. For the factor containing Erfc, we have
\begin{align}\label{eq:CGHS_local_limit_dilaton}
\lim_{\ell\rightarrow0}\ \text{Erfc}\left(\lambda\ell+\frac{b-r}{2\ell}\right)=2\Theta(r-b)\ .
\end{align}
We therefore conclude that equation \eqref{eq:CGHS_non_local_dilaton_solution} reduces to \eqref{eq:CGHS_solved_dilaton} in the local limit $\ell\to0$.

\begin{figure}
  \includegraphics[width=1\columnwidth]{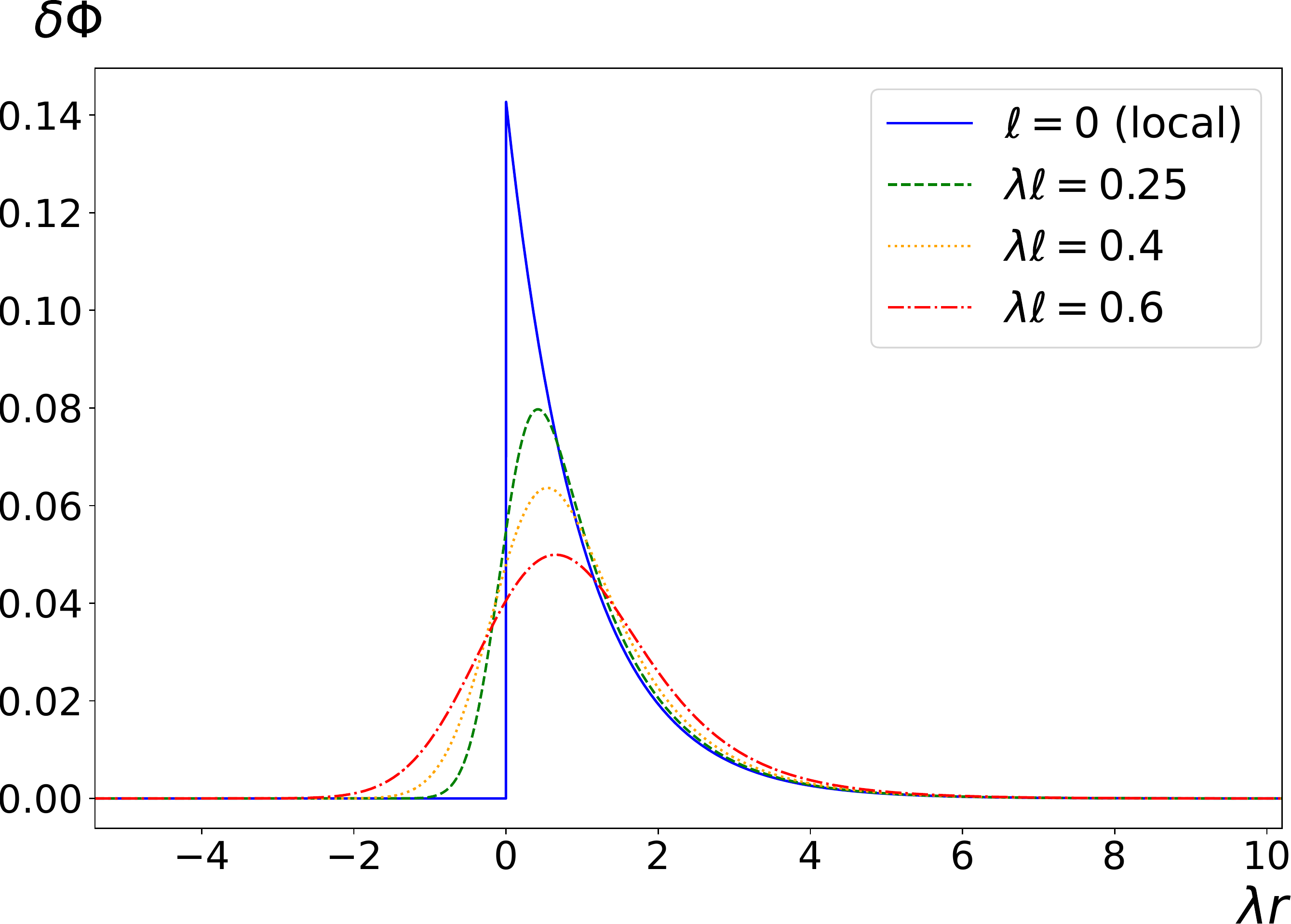}
\caption{Radial dependence of the perturbed dilaton field $\delta\Phi$ for the local and non-local CGHS cases. The solid blue curve corresponds to the local case ($\ell=0$), the dashed green curve corresponds to $\lambda\ell=0.25$, the dotted orange curve corresponds to $\lambda\ell=0.4$ and the red dash-dotted curve corresponds to $\lambda\ell=0.6$. For these plots, we have set $M=10$ and $\lambda=35$.}
\label{fig:delta_Phi_CGHS}
\end{figure}

\begin{figure}
  \includegraphics[width=1\columnwidth]{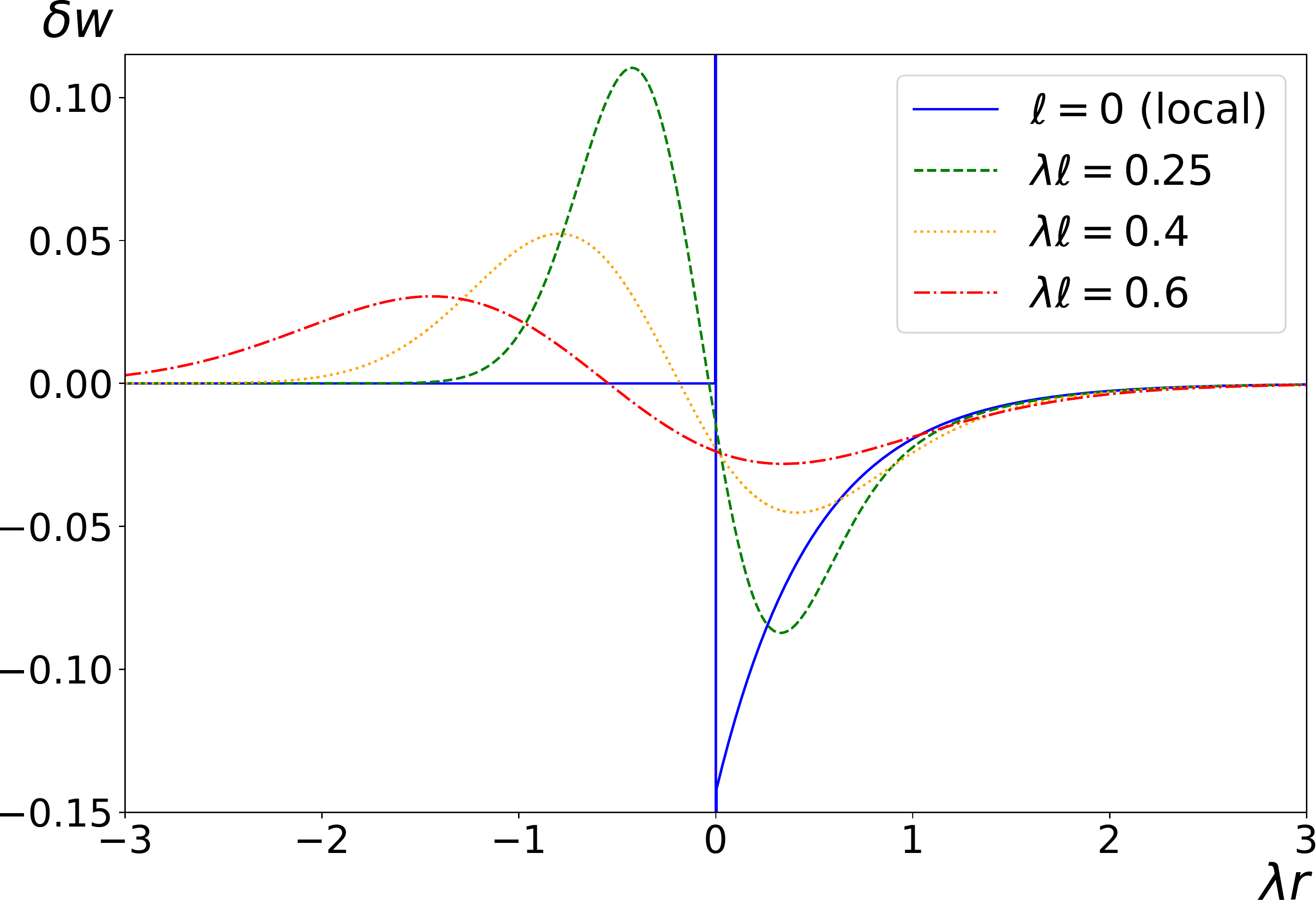}
\caption{Radial dependence of the perturbed conformal scalar $\delta w$ for both the local and non-local CGHS cases. The solid blue curve corresponds to the local case ($\ell=0$), whereas the dashed green, dotted orange and red dash-dotted curves correspond to
$\lambda\ell=0.25,\,0.4$ and $0.6$ respectively. For the local case, the vertical line through the origin is included to illustrate the distributional character of $\delta w$ as well as how it is regularised in the non-local cases. For these plots, we have set $M=10$ and $\lambda=35$.}
\label{fig:delta_w_CGHS}
\end{figure}

\begin{figure}
  \includegraphics[width=1\columnwidth]{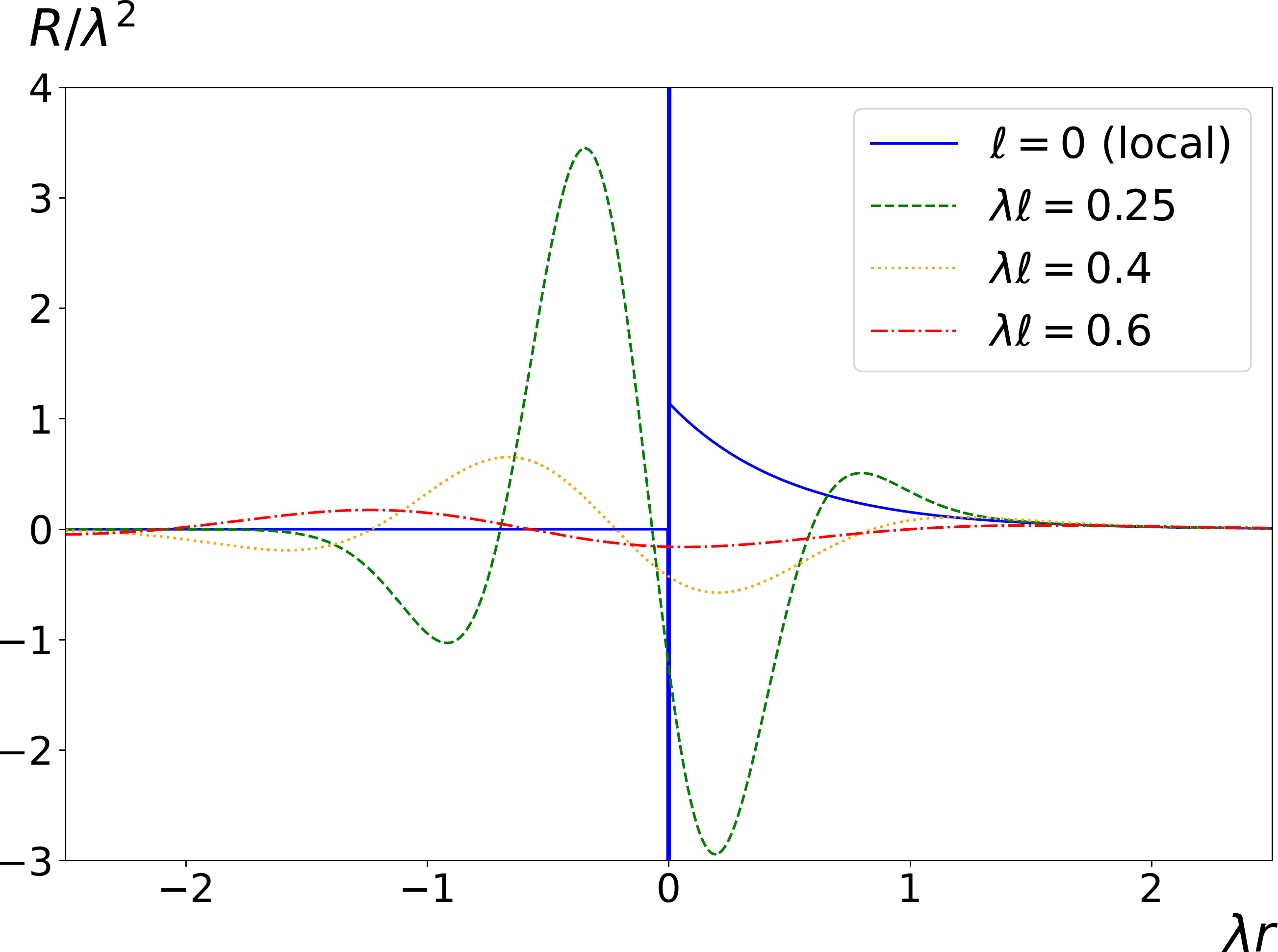}
\caption{Radial dependence of the Ricci scalar $R$ for both the local and non-local CGHS cases when computed up to first order, i.e., $R=-2\Delta\delta w+\mathcal{O}\left(\delta w^2\right)$. The solid blue curve corresponds to the local case ($\ell=0$), whereas the dashed green, dotted orange and red dash-dotted curves correspond to
$\lambda\ell=0.25,\,0.4$ and $0.6$ respectively. For the local case, the vertical line through the origin is included to illustrate the distributional character of the Ricci scalar as well as how it is regularised in the non-local cases. For these plots, we have set $M=10$ and $\lambda=35$.}
\label{fig:R_CGHS}
\end{figure}

Substituting equation \eqref{eq:CGHS_non_local_dilaton_solution} into the constraint equation \eqref{eq:constraint_equation_delta_w} gives the perturbed conformal scalar
\begin{align}\label{eq:CGHS_non_local_metric_solution}
\delta w=-&\frac{M{\text e}^{\ell^2\lambda^2-2\lambda r}}{4\lambda}\text{Erfc}\left(\lambda\ell+\frac{b-r}{2\ell}\right)\nonumber\\
&+\frac{M\left(\lambda\ell+\frac{b-r}{2\ell}\right){\text e}^{-\lambda(r+b)-\frac{(r-b)^2}{4\ell^2}}}{16\sqrt\pi\lambda^3\ell^2}\ .
\end{align}
In Figures \ref{fig:delta_Phi_CGHS}, \ref{fig:delta_w_CGHS} and \ref{fig:R_CGHS} we plot $\delta\Phi$, $\delta w$ and $R/\lambda^2$ respectively with parameter values: $b=0$, $\lambda=35$ and $M=10$. We have chosen these parameter values to illustrate the resolution of the singularity at $r=b$ as well as how the non-local solutions approach the local solution as the length scale of non-locality $\ell$ decreases.

We now wish to examine the effect of non-locality on the linearised CGHS BH solution by taking into account the smallness condition \eqref{eq:smallness_general}. By considering $r\gg b$ and taking $b$ to be large and negative in equations \eqref{eq:CGHS_non_local_dilaton_solution} and \eqref{eq:CGHS_non_local_metric_solution} we obtain
\begin{align}\label{eq:CGHS_approx_delta_Phi_non_local}
\delta\Phi\approx\frac{M{\text e}^{\ell^2\lambda^2-\lambda r}}{2\lambda}\ ,
\end{align}
and
\begin{align}\label{eq:CGHS_approx_delta_w_non_local}
\delta w\approx-\frac{M{\text e}^{\ell^2\lambda^2-2\lambda r}}{2\lambda}\ ,
\end{align}
respectively. For these non-local solutions, the smallness conditions are satisfied provided $r\gg\ell^2\lambda^2+\frac{1}{2\lambda}\ln\frac{M}{\lambda}$. In summary, from equations \eqref{eq:CGHS_approx_delta_Phi_non_local} and \eqref{eq:CGHS_approx_delta_w_non_local} we have found that, in a non-local CGHS theory with $a(\Box)={\text e}^{-\ell^2\Box}$, the CGHS BH mass $M$ is modified by a multiplicative constant factor ${\text e}^{\ell^2\lambda^2}$. This factor, however, can be absorbed through an appropriate coordinate transformation. It is once again straightforward to verify that the conformal gauge constraint equations \eqref{eq:conformal_gauge_constraint_equation_CGHS_v2} and \eqref{eq:conformal_gauge_constraint_equation_2_CGHS_v2} are satisfied at first order when the smallness conditions are taken into account since the non-local solution differs from the local solution by the aforesaid multiplicative factor in such a case.


\section{Conclusions}
\label{Conclusions}
In this paper, we constructed ghost-free infinite-derivative modifications for the SRG and CGHS dilaton gravity theories. For the SRG theory, we assumed the Schwarzschild-type gauge and diagonalised the quadratic action which contains two off-shell degrees of freedom. We constructed a source action that, upon taking into account the smallness conditions, could be used to generate the linearised spherically-reduced Schwarzschild solution of the local theory. Inspired by the diagonalised quadratic action in the local theory, we constructed ghost-free infinite-derivative modifications of the SRG theory. By taking the two operators containing infinitely many derivatives to be the exponential operator ${\text e}^{-\ell^2\Box}$ we were able to obtain a non-local modification of the spherically-reduced Schwarzschild solution after including the same source action used in the local case. We found that, in the context of this ghost-free infinite-derivative SRG theory, the $1/r$ factor in the linearised metric is weighted by the error function $\text{Erf}\left(r/2\ell\right)$; resolving the singular nature of the local linearised solution at $r=0$. In \cite{Biswas:2011ar}, it was found that the $1/r$ nature in the linearised Schwarzschild solution of GR is also resolved through the error function in four-dimensional IDG.

In the case of the local CGHS theory, we specified the conformal gauge and studied perturbations around a general background solution. We diagonalised the CGHS quadratic action and isolated its one propagating off-shell degree of freedom. The ghost-free infinite-derivative modification of the CGHS quadratic action then involved the inclusion of one non-zero analytic differential operator containing infinitely many derivatives. We made use of the source action \eqref{eq:source_action_CGHS_for_conclusion} to generate a solution in the local theory which coincides with the linearised CGHS BH solution when taking the smallness conditions into account. The obtained solution is also singular at some given position $b$. Nonetheless, in the non-local theory with form factor ${\text e}^{-\ell^2\Box}$, we found that the solution is weighted by the complementary error function $\text{Erfc}\left(\lambda\ell+(b-r)/2\ell\right)$ which allowed for the singular nature appearing in the local solution to be resolved.

While we have only considered ghost-free infinite-derivative modifications of the SRG and CGHS gravity theories, there is still a multitude of two-dimensional dilaton gravity theories (discussed extensively in \cite{Grumiller:2002nm}) for which similar modifications as the ones presented here can be constructed. One example would be to construct ghost-free infinite-derivative modifications of the dilaton action obtained through the spherical reduction of GR in dimensions other than four. Another example would be to study ghost-free infinite-derivative modifications of the CGHS theory containing additional scalar matter fields such as the model considered in \cite{Callan:1992rs}. In this context, one could investigate how solutions generated through so-called $f$-waves are modified as a result of introducing non-locality. There are also two-dimensional dilaton gravity models that include fermionic matter \cite{Pelzer:1998ea,Cavaglia:1998yp} for which one could investigate ghost-free infinite-derivative modifications.

We also note that we were unable to obtain full solutions to these infinite-derivative dilaton gravity theories and were only able to examine how linearised local solutions were modified in the non-local theories. Thus, there is still the question of whether one can find full non-local solutions to ghost-free infinite-derivative dilaton gravity in two space-time dimensions.

\appendix

\section{Derivation of equation \texorpdfstring{\eqref{eq:quadratic_action_general}}{}}\label{sec:derivation_quadratic}
Here, we provide the main steps to obtaining the quadratic action associated with the general dilaton model \eqref{eq:action_with_Phi} without any gauge fixing or the specification of a particular background solution. We begin by first obtaining the necessary second-order functional derivatives. To this end, we vary equation \eqref{eq:first_variation_dilaton} and write
\begin{align}\label{eq:vary_S_Phi}
&\delta\frac{\delta S_{\text{local}}}{\delta\Phi}=-2\sqrt{-g}g_{\mu\nu}\delta g^{\mu\nu}\left[\frac12\Phi R-2k\Box\Phi+2\Phi\lambda^2\right]\nonumber\\
&+4\sqrt{-g}\bigg[\frac12R\delta\Phi+\frac{\Phi}{2}\big(R_{\mu\nu}\delta g^{\mu\nu}-\nabla_\mu \nabla_\nu\delta g^{\mu\nu}\nonumber\\
&+g_{\mu\nu}\Box\delta g^{\mu\nu}\big)-2k\delta(\Box)\Phi-2k\Box\delta\Phi+2\delta\Phi\lambda^2\bigg]\ .
\end{align}
We can simplify the right-hand-side of this expression by using the fact that the Einstein tensor is zero in two space-time dimensions. In addition, we can evaluate the $\delta\left(\Box\right)$ term by using the expression \cite{Biswas:2013cha}
\begin{align}\label{eq:var_box}
\delta(\Box)\Phi=\delta g^{\mu\nu}\nabla_\mu \nabla_\nu\Phi+&\nabla_\mu\Phi \nabla_\nu\delta g^{\mu\nu}\nonumber\\
&-\frac12g_{\mu\nu}\nabla_\alpha\Phi \nabla^\alpha\delta g^{\mu\nu}\ ,
\end{align}
which follows directly from
\begin{align}\label{eq:vary_D_D}
\delta\left(\nabla_\mu \nabla_\nu\right)&\Phi=\partial_\alpha \Phi\bigg[g_{\sigma(\nu}\nabla_{\mu)}\delta g^{\alpha\sigma}
-\frac12g_{\mu\rho}g_{\nu\beta}\nabla^\alpha\delta g^{\rho\beta}\bigg]\ .
\end{align}
We can now write equation \eqref{eq:vary_S_Phi} as
\begin{align}\label{eq:vary_var_Phi}
&\delta\frac{\delta S_{\text{local}}}{\delta\Phi}=-4\sqrt{-g}\ g_{\mu\nu}\delta g^{\mu\nu}\left[\Phi\lambda^2-k\Box\Phi\right]\nonumber\\
&+4\sqrt{-g}\bigg[\frac12R\delta\Phi-\frac12\Phi \nabla_\mu \nabla_\nu\delta g^{\mu\nu}+\frac12\Phi g_{\mu\nu}\Box\delta g^{\mu\nu}\nonumber\\
&-2k\delta g^{\mu\nu}\nabla_\mu \nabla_\nu\Phi-2k\nabla_\mu\Phi \nabla_\nu\delta g^{\mu\nu}\nonumber\\
&+kg_{\mu\nu}\nabla_\alpha\Phi \nabla^\alpha\delta g^{\mu\nu}-2k\Box\delta\Phi+2\delta\Phi\lambda^2\bigg]\ .
\end{align}
In this form, we can now use equation \eqref{eq:vary_var_Phi} to extract the second order functional derivatives
\begin{align}\label{eq:vary_Phi_Phi}
\frac{\delta^2S_{\text{local}}}{\delta\Phi(x')\delta\Phi(x)}&=4\sqrt{-g}\bigg[\frac12R-2k\Box+2\lambda^2\bigg]\nonumber\\
&\times\delta^{(2)}(x-x')\ ,
\end{align}
and
\begin{align}\label{eq:vary_g_Phi}
&\frac{\delta^2S_{\text{local}}}{\delta g^{\mu\nu}(x')\delta\Phi(x)}=4\sqrt{-g}\bigg\{-g_{\mu\nu}\left[\Phi\lambda^2-k\Box\Phi\right]\nonumber\\
&+\frac12\Phi g_{\mu\nu}\Box-\frac12\Phi \nabla_{\mu}\nabla_{\nu}-2k\nabla_\mu \nabla_\nu\Phi\nonumber\\
&-2k\nabla_\mu\Phi \nabla_\nu+kg_{\mu\nu}\nabla_\alpha\Phi \nabla^\alpha\bigg\}\delta^{(2)}(x-x')\ ,
\end{align}
where $\delta^{(2)}(x-x')$ is the two-dimensional Dirac delta function. It now remains to derive the two-remaining second order functional derivatives. Through the variation of equation \eqref{eq:first_variation_metric} one obtains
\begin{align}\label{eq:vary_var_g}
&\frac{1}{\sqrt{-g}}\delta\frac{\delta S_{\text{local}}}{\delta g^{\mu\nu}}=\delta g^{\alpha\beta}\bigg\{\big[g_{\alpha\beta}g_{\mu\nu}+2g_{\mu\alpha}g_{\nu\beta}\big]\nonumber\\
&\times\big[k\left(\partial\Phi\right)^2+\Phi^2\lambda^2+\frac{a^2}{2}\big]
-\frac12g_{\alpha\beta}\big[g_{\mu\nu}\Box\Phi^2-\nabla_\mu \nabla_\nu\Phi^2\nonumber\\
&+4k\partial_\mu\Phi\partial_\nu\Phi\big]+g_{\mu\nu}\big[\nabla_\alpha \nabla_\beta\Phi^2-2k\partial_\alpha\Phi\partial_\beta\Phi\big]\nonumber\\
&-g_{\mu\alpha}g_{\nu\beta}\Box\Phi^2\bigg\}+\nabla_\sigma\delta g^{\alpha\beta}\bigg[g_{\mu\nu}\delta^\sigma_\beta \nabla_\alpha\Phi^2\nonumber\\
&-g_{\mu[\nu}g_{\alpha]\beta}\nabla^\sigma\Phi^2-g_{\beta(\nu}\delta^\sigma_{\mu)} \nabla_\alpha\Phi^2\bigg]\nonumber\\
&+2\Big[g_{\mu\nu}\Box\left(\Phi\delta\Phi\right)-\nabla_\mu \nabla_\nu\left(\Phi\delta\Phi\right)+4k\partial_{(\mu}\Phi\partial_{\nu)}\delta\Phi\nonumber\\
&-2kg_{\mu\nu}\partial_\alpha\Phi\partial^\alpha\delta\Phi-2g_{\mu\nu}\Phi\delta\Phi\lambda^2\Big]\ .
\end{align}
Equation \eqref{eq:vary_var_g} can be used to extract
\begin{align}\label{eq:vary_Phi_g}
&\frac{\delta^2S_{\text{local}}}{\delta\Phi(x')\delta g^{\mu\nu}(x)}=2\sqrt{-g}\bigg[\Big(g_{\mu\nu}\Box-\nabla_\mu \nabla_\nu\Big)\nonumber\\
&\times\left(\Phi\delta^{(2)}(x-x')\right)+\Big(4k\partial_{(\mu}\Phi\partial_{\nu)}-2kg_{\mu\nu}\partial_\alpha\Phi\partial^\alpha\nonumber\\
&-2g_{\mu\nu}\Phi\lambda^2\Big)\delta^{(2)}(x-x')
\bigg]\ ,
\end{align}
and
\begin{align}\label{eq:vary_g_g}
&\frac{\delta^2S_{\text{local}}}{\delta g^{\alpha\beta}(x')\delta g^{\mu\nu}(x)}=\sqrt{-g}\bigg\{\left[g_{\mu\nu}g_{\alpha\beta}+2g_{\mu\alpha}g_{\nu\beta}\right]\nonumber\\
&\times\left[k\left(\partial\Phi\right)^2+\Phi^2\lambda^2+\frac{a^2}{2}\right]-\frac12g_{\alpha\beta}\bigg[g_{\mu\nu}\Box\Phi^2-\nabla_\mu \nabla_\nu\Phi^2\nonumber\\
&+4k\partial_\mu\Phi\partial_\nu\Phi\bigg]+g_{\mu\nu}\Big[\nabla_\alpha \nabla_\beta\Phi^2-2k\partial_\alpha\Phi\partial_\beta\Phi\Big]\nonumber\\
&-g_{\mu\alpha}g_{\nu\beta}\Box\Phi^2+\Big[g_{\mu\nu}\delta^\sigma_\beta \nabla_\alpha\Phi^2-g_{\mu[\nu}g_{\alpha]\beta}\nabla^\sigma\Phi^2\nonumber\\
&-g_{\beta(\mu}\delta^\sigma_{\nu)}\nabla_\alpha\Phi^2\Big]\nabla_\sigma
\bigg\}\delta^{(2)}(x-x')\ .
\end{align}
The desired quadratic action can be found by substituting equations \eqref{eq:vary_Phi_Phi}, \eqref{eq:vary_g_Phi}, \eqref{eq:vary_Phi_g} and \eqref{eq:vary_g_g} into the definition \eqref{eq:quadratic_action_definition}. In order to obtain this, let us write the quadratic action as
\begin{align}\label{eq:quadratic_action_I_i}
\delta^2S_{\text{local}}=\sum_{i=1}^4I_i\ ,
\end{align}
where we define
\begin{align}\label{eq:I_1}
I_1:=\int\text{d}^2x\ \text{d}^2x'\delta\Phi(x)\delta\Phi(x')\frac{\delta^2S_{\text{local}}}{\delta\Phi(x)\delta\Phi(x')}\Bigg|_{\left(\bar g_{\mu\nu},\bar\Phi\right)}\ ,
\end{align}
\begin{align}\label{eq:I_2}
I_2:=\int\text{d}^2x\ \text{d}^2x'\delta\Phi(x)\delta g^{\mu\nu}(x')\frac{\delta^2S_{\text{local}}}{\delta\Phi(x)\delta g^{\mu\nu}(x')}\Bigg|_{\left(\bar g_{\mu\nu},\bar\Phi\right)}\ ,
\end{align}
\begin{align}\label{eq:I_3}
I_3:=\int\text{d}^2x\ \text{d}^2x'\delta g^{\mu\nu}(x)\delta\Phi(x')\frac{\delta^2S_{\text{local}}}{\delta g^{\mu\nu}(x)\delta\Phi(x')}\Bigg|_{\left(\bar g_{\mu\nu},\bar\Phi\right)}\ ,
\end{align}
and
\begin{align}\label{eq:I_4}
I_4:=\int\text{d}^2x\ \text{d}^2x'\delta g^{\mu\nu}(x)\delta g^{\alpha\beta}(x')\frac{\delta^2S_{\text{local}}}{\delta g^{\mu\nu}(x)\delta g^{\alpha\beta}(x')}\Bigg|_{\left(\bar g_{\mu\nu},\bar\Phi\right)}\ .
\end{align}
Let us now consider each of the $I_i$ individually. Evaluating equation \eqref{eq:vary_Phi_Phi} at the background solution $\left(\bar g_{\mu\nu},\bar\Phi\right)$ and substituting this into equation \eqref{eq:I_1} gives
\begin{align}\label{eq:I_1_2}
I_1=4\int\text{d}^2x\sqrt{-\bar g}\ \delta\Phi\left(\frac12\bar R-2k\bar\Box+2\lambda^2\right)\delta\Phi\ ,
\end{align}
where $\bar R$ and $\bar\Box:={\bar g}^{\mu\nu}\bar \nabla_\mu\bar \nabla_\mu$ are the background Ricci scalar and d'Alembertian respectively. The equation of motion resulting from \eqref{eq:first_variation_dilaton} tells us that
\begin{align}\label{eq:appendix_eom}
\frac12\bar R+2\lambda^2=2k\frac{\bar\Box\bar\Phi}{\bar\Phi}\ ,
\end{align}
and thus, equation \eqref{eq:I_1_2} becomes
\begin{align}\label{eq:I_1_3}
I_1=8k\int\text{d}^2x\sqrt{-\bar g}\ \delta\Phi\left(\frac{\bar\Box\bar\Phi}{\bar\Phi}-\bar\Box\right)\delta\Phi\ .
\end{align}
We now turn our attention to finding $I_2$. By evaluating \eqref{eq:vary_g_Phi} at the background solution, substituting this into equation \eqref{eq:I_2} and again using the equation of motion \eqref{eq:appendix_eom}, one finds
\begin{align}\label{eq:I_2_2}
I_2=4\int\text{d}^2x\sqrt{-\bar g}\ \delta g^{\mu\nu}\bigg[&\frac14\bar R\bar g_{\mu\nu}\bar\Phi\delta\Phi+\frac12\bar g_{\mu\nu}\bar\Box\left(\bar\Phi\delta\Phi\right)\nonumber\\
-\frac12\bar \nabla_\mu\bar \nabla_\nu\left(\bar\Phi\delta\Phi\right)&+2k\partial_\mu\bar\Phi\partial_\nu\delta\Phi\nonumber\\
&-k\bar g_{\mu\nu}\bar \nabla_\alpha\left(\delta\Phi\partial^\alpha\bar\Phi\right)\bigg]\ .
\end{align}
Similarly, by evaluating equation \eqref{eq:vary_Phi_g} at the background solution and substituting this into \eqref{eq:I_3} we obtain
\begin{align}\label{eq:I_3_2}
&I_3=4\int\text{d}^2x\sqrt{-\bar g}\ \delta g^{\mu\nu}\bigg[\frac12\Big(\bar g_{\mu\nu}\bar\Box-\bar \nabla_\mu\bar \nabla_\nu\Big)\left(\bar\Phi\delta\Phi\right)\nonumber\\
&+2k\partial_{\mu}\bar\Phi\partial_{\nu}\delta\Phi-\bar g_{\mu\nu}\bar \nabla_\alpha\left(\delta\Phi\partial^\alpha\bar\Phi\right)+(1-k)\bar g_{\mu\nu}\delta\Phi\bar\Box\bar\Phi\nonumber\\
&+\frac14\bar R\bar g_{\mu\nu}\bar\Phi\delta\Phi\bigg]\ ,
\end{align}
where we have once again made use of the equation of motion \eqref{eq:appendix_eom}.

It now remains to compute $I_4$. By considering the second order functional derivative \eqref{eq:vary_g_g} at the classical solution and substituting the result into equation \eqref{eq:I_4} one finds the integral
\begin{align}\label{eq:I_4_2}
&I_4=\int\text{d}^2x\sqrt{-\bar g}\ \delta g^{\mu\nu}\bigg\{\bar g_{\alpha\beta}\Big[\bar g_{\mu\nu}\Big(k\left(\partial\bar\Phi\right)^2+\bar\Phi^2\lambda^2+\frac{a^2}{2}\Big)\nonumber\\
&-\frac12\Big(\bar g_{\mu\nu}\bar\Box\bar\Phi^2-\bar \nabla_\mu\bar \nabla_\nu\bar\Phi^2+4k\partial_\mu\bar\Phi\partial_\nu\bar\Phi\Big)\Big]\nonumber\\
&+\bar g_{\mu\alpha}\Big[2\bar g_{\nu\beta}\Big(k\left(\partial\bar\Phi\right)^2+\bar\Phi^2\lambda^2+\frac{a^2}{2}\Big)-\bar g_{\nu\beta}\bar\Box\bar\Phi^2\Big]\nonumber\\
&+\Big[\bar g_{\mu\nu}\delta^\sigma_\beta\bar \nabla_\alpha\bar\Phi^2-\bar g_{\mu[\nu}\bar g_{\alpha]\beta}\bar \nabla^\sigma\bar\Phi^2\nonumber\\
&\ \ \ \ \ \ \ \ \ \ \ \ \ \ \ \ \ \ \ \ \ \ \ \ \ \ \ -\bar g_{\beta\mu}\delta^\sigma_{\nu}\bar \nabla_\alpha\bar\Phi^2\Big]\bar \nabla_\sigma
\bigg\}\delta g^{\alpha\beta}\ .
\end{align}
The integral above can be simplified significantly by leveraging the equation of motion obtained by setting \eqref{eq:first_variation_metric} to zero. More specifically, using this equation of motion, the first square bracket in the integral \eqref{eq:I_4_2} vanishes while the second bracket can be simplified. This allows us to write equation \eqref{eq:I_4_2} as
\begin{align}\label{eq:I_4_3}
I_4=\int\text{d}^2x\sqrt{-\bar g}\ \delta g^{\mu\nu}\bigg\{&\bar g_{\mu\alpha}\Big[4k\partial_\beta\bar\Phi\partial_\nu\bar\Phi-\bar \nabla_\nu\bar \nabla_\beta\bar\Phi^2\Big]\nonumber\\
+\bar g_{\mu\nu}\bar \nabla_\alpha\bar\Phi^2&\bar \nabla_\beta-\bar g_{\mu[\nu}\bar g_{\alpha]\beta}\bar \nabla^\sigma\bar\Phi^2\bar \nabla_\sigma\nonumber\\
&-\bar g_{\beta\mu}\bar \nabla_\alpha\bar\Phi^2\bar \nabla_{\nu}
\bigg\}\delta g^{\alpha\beta}\ .
\end{align}
Substituting equations \eqref{eq:I_1_3}, \eqref{eq:I_2_2}, \eqref{eq:I_3_2} and \eqref{eq:I_4_3} into \eqref{eq:quadratic_action_I_i} gives the desired quadratic action \eqref{eq:quadratic_action_general}.

\section{Dilaton gravity in the conformal and Schwarzschild-type gauges}
\subsection{Dilaton gravity in the conformal gauge}\label{sec:dilaton_gravity_in_conformal_gauge}
In this section, we wish to show that the action
\begin{align}\label{eq:dilaton_action_gauge_fixed_general}
S_{\text{local}}=4\int\text{d}^2x\bigg[\Phi^2{\text{e}}^{2w}&\lambda^2+\frac{a^2}{2}{\text{e}}^{2w}+k\eta^{\mu\nu}\partial_\mu\Phi\partial_\nu\Phi\nonumber\\
&-\frac12\Phi^2\eta^{\mu\nu}\partial_\mu\partial_\nu w\bigg]\ ,
\end{align}
which is obtained by specifying the conformal gauge $g_{\mu\nu}={\text{e}}^{2w}\eta_{\mu\nu}$ in equation \eqref{eq:action_with_Phi} where $\eta_{\mu\nu}=\text{diag}(-1,1)$ is the Minkowski metric in Cartesian coordinates contains all the dynamics of the original action \eqref{eq:action_with_Phi}. We wish to show that the same dynamical equations of motion are obtained irrespective of whether we specify the conformal gauge at the level of the action or at the level of the field equations. Such a check is necessary since the two approaches are not equivalent in general. 

Let us begin by stating the field equations associated with the gauge-fixed action \eqref{eq:dilaton_action_gauge_fixed_general}. Variation with respect to the dilaton field gives
\begin{align}\label{eq:dilaton_eom_gauge_at_action}
\frac{\delta S_{\text{local}}}{\delta\Phi}=8\Big[\Phi {\text{e}}^{2w}&\lambda^2-k\eta^{\mu\nu}\partial_\mu\partial_\nu\Phi\nonumber\\
&-\frac12\Phi\eta^{\mu\nu}\partial_\mu\partial_\nu w\Big]\ ,
\end{align}
while variation with respect to the conformal scalar $w$ gives
\begin{align}\label{eq:metric_eom_gauge_at_action}
\frac{\delta S_{\text{local}}}{\delta w}=4\Big[2\Phi^2{\text{e}}^{2w}&\lambda^2+2\frac{a^2}{2}{\text{e}}^{2w}\nonumber\\
&-\frac12\eta^{\mu\nu}\partial_\mu\partial_\nu\Phi^2\Big]\ .
\end{align}
We now turn our attention to specifying the conformal gauge in equations \eqref{eq:first_variation_dilaton} and \eqref{eq:first_variation_metric_trace}.

Let us first study the action of the space-time covariant d'Alembertian operator $\Box:=\nabla_\mu \nabla^\mu$ on some smooth test function $h$. Written in terms of the connection coefficients $\Gamma^\alpha_{\ \mu\nu}$ we have
\begin{align}
\Box h=g^{\mu\nu}\partial_\mu\partial_\nu h-g^{\mu\nu}\Gamma^\alpha_{\ \mu\nu}\partial_\alpha h\ .
\end{align}
In the conformal gauge, we have
\begin{align}
g^{\mu\nu}\Gamma^\alpha_{\ \mu\nu}={\text{e}}^{-2w}\eta^{\alpha\sigma}\big(2-\eta_{\mu\nu}\eta^{\mu\nu}\big)\partial_\sigma w=0\ .
\end{align}
It follows that
\begin{align}\label{eq:Box_h_conformal_gauge}
\Box h={\text{e}}^{-2w}\eta^{\mu\nu}\partial_\mu\partial_\nu h\ .
\end{align}
By specifying the conformal gauge in equation \eqref{eq:first_variation_dilaton} one obtains
\begin{align}\label{eq:dilaton_eom_gauge}
\frac{\delta S_{\text{local}}}{\delta\Phi}=4&{\text{e}}^{2w}\big[-\Phi {\text{e}}^{-2w}\eta^{\mu\nu}\partial_\mu\partial_\nu w\nonumber\\
&-2k{\text{e}}^{-2w}\eta^{\mu\nu}\partial_\mu\partial_\nu\Phi+2\Phi\lambda^2\big]\ ,
\end{align}
where we made use of equations \eqref{eq:Ricci_scalar_conformal_gauge} and \eqref{eq:Box_h_conformal_gauge}. Upon simplification, it is readily seen that equation \eqref{eq:dilaton_eom_gauge} is nothing more than equation \eqref{eq:dilaton_eom_gauge_at_action}. Fixing the conformal gauge in equation \eqref{eq:first_variation_metric_trace} gives
\begin{align}\label{eq:first_variation_metric_trace_gauge_fixed}
-2g^{\mu\nu}\frac{\delta S_{\text{local}}}{\delta g^{\mu\nu}}=-2{\text{e}}^{2w}&\bigg[{\text{e}}^{-2w}\eta^{\mu\nu}\partial_\mu\partial_\nu\Phi^2\nonumber\\
&-4\left(\Phi^2\lambda^2+\frac{a^2}{2}\right)\bigg]\ .
\end{align}
Simplifying the last result shows that the right-hand-side of equation \eqref{eq:first_variation_metric_trace_gauge_fixed} is nothing more than equation \eqref{eq:metric_eom_gauge_at_action}.

Finally, notice that in this case the equivalence with the trace of the field equations is sufficient to prove the validity of the gauge-fixing approach, since such equation contains all the dynamics to obtain the solution. On the other hand, the trace-free part of the field equations provides us with the constraint equations
\begin{align}\label{eq:conformal_gauge_constraint_equation_v2}
\delta^{\mu\nu}\left(4k\partial_\mu\Phi\partial_\nu\Phi+2\partial_{\mu}w\partial_{\nu}\Phi^2-\partial_\mu\partial_\nu\Phi^2\right)=0\ ,
\end{align}
and
\begin{align}\label{eq:conformal_gauge_constraint_equation_2_v2}
4k\partial_\mu\Phi\partial_\nu\Phi+2\partial_{(\mu}w\partial_{\nu)}\Phi^2-\partial_\mu\partial_\nu\Phi^2=0\ ,\ \ \ \ (\mu\neq\nu)\ ,
\end{align}
associated with the imposing of the conformal gauge. We therefore conclude that the consideration of the gauge-fixed action \eqref{eq:dilaton_action_gauge_fixed_general} together with the constraint equations \eqref{eq:conformal_gauge_constraint_equation_v2} and \eqref{eq:conformal_gauge_constraint_equation_2_v2} above is equivalent to the consideration of the original action \eqref{eq:action_with_Phi}.

\subsection{Dilaton gravity in the Schwarzschild-type gauge}\label{sec:dilaton_gravity_in_Schwarzschild_gauge}

In this section, we wish to show that the action \eqref{eq:action_with_Phi} gives the same dynamical equations of motion regardless of whether we specify the Schwarzschild-type gauge at the level of the action or at the level of the field equations. The Schwarzschild-type gauge is stated in equation \eqref{eq:Schwarzschild_gauge}. Given this form of the metric, the non-vanishing components of the connection coefficients are
\begin{align}\label{eq:dot_Christoffel_symbols}
\Gamma^t_{\ tt}=-\Gamma^r_{\ rt}=-f^2\Gamma^t_{\ rr}=\frac{\dot f}{2f}\ ,
\end{align}
and
\begin{align}\label{eq:prime_Christoffel_symbols}
\Gamma^t_{\ rt}=-\Gamma^r_{\ rr}=\frac{1}{f^2}\Gamma^r_{\ tt}=\frac{f'}{2f}\ ,
\end{align}
where $f'$ and $\dot f$ denote differentiation with respect to $r$ and $t$ respectively. From the above we can compute the non-vanishing Ricci tensor components
\begin{align}
R_{tt}=\frac{ff''}2+\frac{\ddot f}{2f}-\frac{\dot f^2}{f^2}\ ,
\end{align}
and
\begin{align}
R_{rr}=-\frac{\ddot f}{2f^3}+\frac{\dot f^2}{f^4}-\frac{f''}{2f}\ .
\end{align}
From the last two expressions it follows that the Ricci scalar is
\begin{align}\label{eq:Ricci_scalar_schwarzschild_gauge}
R=-\frac{\ddot f}{f^2}-f''+\frac{2\dot f^2}{f^3}\ .
\end{align}
By specifying the Schwarzschild-type gauge in the action \eqref{eq:action_with_Phi} we have
\begin{align}\label{eq:schwarzschild_fix_gauge_total_v2}
S_{\text{local}}=\int\text{d}^2x\bigg[R\Phi^2-\frac{4k}{f}\dot\Phi^2+4kf\Phi'^2+4\Phi^2\lambda^2+2a^2\bigg]\ ,
\end{align}
and it is understood that the Ricci scalar is given by equation \eqref{eq:Ricci_scalar_schwarzschild_gauge}. The variation of this gauge-fixed action yields
\begin{align}\label{eq:SRG_appendix_vary}
\delta&S_{\text{local}}=\int\text{d}^2x\bigg[\delta R\Phi^2+2R\Phi\delta\Phi+\frac{4k\delta f}{f^2}\dot\Phi^2\nonumber\\
&+4k\delta f\Phi'^2+8kg^{\mu\nu}\partial_\mu\Phi\partial_\nu\delta\Phi+8\Phi\delta\Phi\lambda^2\bigg]\ .
\end{align}
Let us turn our attention to the variation of the Ricci scalar $\delta R$. From equation \eqref{eq:Ricci_scalar_schwarzschild_gauge} we have
\begin{align}
\delta R=-\partial_t^2\left(\frac{\delta f}{f^2}\right)-\partial_r^2\delta f\ .
\end{align}
In terms of the perturbed inverse metric $\delta g^{\mu\nu}$ the last expression is nothing more than
\begin{align}
\delta R=-\partial_\mu\partial_\nu\delta g^{\mu\nu}\ .
\end{align}
We can now write equation \eqref{eq:SRG_appendix_vary} as
\begin{align}
&\delta S_{\text{local}}=2\int\text{d}^2x\bigg\{\delta\Phi\left(R+4k\nabla^2+4\lambda^2\right)\Phi +\delta f \nonumber\\
&\times\bigg[\frac{(2k-1)\dot\Phi^2}{f^2}+(2k-1)\Phi'^2-\frac{\Phi\ddot\Phi}{f^2}-\Phi\Phi''\bigg]
\bigg\}\ .
\end{align}
The first functional derivatives of the gauge-fixed action with respect to $\Phi$ and $f$ are now found to be
\begin{align}\label{eq:Phi_eom_schwarzschild_gauge}
\frac{\delta S_{\text{local}}}{\delta\Phi}=2\left(R-4k\nabla^2+4\lambda^2\right)\Phi\ ,
\end{align}
and
\begin{align}\label{eq:f_eom_schwarzschild_gauge}
\frac12\frac{\delta S_{\text{local}}}{\delta f}=\frac{(2k-1)\dot\Phi^2}{f^2}+(2k-1)\Phi'^2-\frac{\Phi\ddot\Phi}{f^2}-\Phi\Phi''\ ,
\end{align}
respectively. We now wish to show that these are the same equations of motion that are obtained when gauge-fixing at the level of the field equations. It is immediately clear that equations \eqref{eq:Phi_eom_schwarzschild_gauge} and \eqref{eq:first_variation_dilaton} coincide. We therefore turn our attention to imposing the Schwarzschild-type gauge in equation \eqref{eq:first_variation_metric}.

It is worth mentioning that the perturbed metric $\delta g^{\mu\nu}$ is traceless in the Schwarzschild-type gauge, i.e., $g_{\mu\nu}\delta g^{\mu\nu}=0$. We therefore examine the trace-free part of equation \eqref{eq:first_variation_metric}
\begin{align}
\left(\frac{\delta S_\text{local}}{\delta g^{\mu\nu}}\right)^{\text{TF}}:=\frac{\delta S_\text{local}}{\delta g^{\mu\nu}}-\frac12g_{\mu\nu}g^{\alpha\beta}\frac{\delta S_\text{local}}{\delta g^{\beta\alpha}}\ ,
\end{align}
where we have used the superscript $\text{TF}$ to denote the trace-free part. From equation \eqref{eq:first_variation_metric} we have the following after imposing the Schwarzschild-type gauge
\begin{align}
\left(\frac{\delta S_\text{local}}{\delta g^{rr}}\right)^{\text{TF}}=2k&\Phi'^2+\frac{2k\dot\Phi^2}{f^2}\nonumber\\
&-\frac{1}{2f^2}\nabla_t\partial_t\Phi^2-\frac12\nabla_r\partial_r\Phi^2\ .
\end{align}
By replacing the Levi-Civita covariant derivatives in the above with partial derivatives and connection coefficients, we have
\begin{align}\label{eq:gauge_fix_eom_with_Christoffel}
\left(\frac{\delta S_\text{local}}{\delta g^{rr}}\right)^{\text{TF}}&=2k\Phi'^2+2k\frac{\dot\Phi^2}{f^2}-\frac{\partial^2_t\Phi^2}{2f^2}-\frac12\partial^2_r\Phi^2\nonumber\\
&+\frac{1}{2f^2}\Gamma^\alpha_{\ tt}\partial_\alpha\Phi^2+\frac12\Gamma^\alpha_{\ rr}\partial_\alpha\Phi^2\ .
\end{align}
From equations \eqref{eq:dot_Christoffel_symbols} and \eqref{eq:prime_Christoffel_symbols} we have
\begin{align}
\Gamma^\alpha_{\ tt}=-f^2\Gamma^\alpha_{\ rr}\ .
\end{align}
Upon substituting this expression into equation \eqref{eq:gauge_fix_eom_with_Christoffel}, we find that the result is nothing more than the right-hand-side of equation \eqref{eq:f_eom_schwarzschild_gauge}.

Finally, using a similar argument as in the previous subsection, the equivalence with respect to the trace-free part is enough to motivate the gauge fixing approach since it contains all the dynamics to obtain the solution. While the action \eqref{eq:schwarzschild_fix_gauge_total_v2} contains all the dynamics, we also have the constraint equations obtained from the trace equation \eqref{eq:first_variation_metric_trace} and the $(t,r)$ component of \eqref{eq:first_variation_metric} with the former providing us with
\begin{align}
f\partial^2_r\Phi^2-\frac{1}{f}\partial_t^2\Phi^2+&f'\partial_r\Phi^2+\frac{\dot f\partial_t\Phi^2}{f^2}\nonumber\\
&-4\Phi^2\lambda^2-2a^2=0\ ,
\end{align}
and the latter yielding
\begin{align}
\partial_t\partial_r\Phi^2-\frac{f'\partial_t\Phi^2}{2f}+\frac{\dot f\partial_r\Phi^2}{2f}-4k\dot\Phi\Phi'=0\ .
\end{align}

\section{Derivation of equation \texorpdfstring{\eqref{eq:quadratic_action_SRG}}{}}\label{sec:derivation_SRG}
Here we derive the quadratic action associated with the SRG theory which is given in equation \eqref{eq:quadratic_action_SRG}. We will obtain this result by specifying the Schwarzschild-type gauge in equation \eqref{eq:quadratic_action_general} for the case of the SRG theory and taking the background fields to be the flat space-time solution with a linear dilaton. We remind the reader that the SRG theory is described by the dilaton model \eqref{eq:action_with_Phi} for the choice of parameters: $\lambda=0$, $k=1/2$ and $a$ left unspecified. We fix the Schwarzschild-type gauge by taking the metric to be of the form given in equation \eqref{eq:Schwarzschild_gauge}. We take the background metric to be $\bar g_{\mu\nu}=\eta_{\mu\nu}$ while taking the background dilaton field to be $\bar\Phi=ar$ which correspond to the flat space-time solution. As already discussed in Section \ref{sec:SRG}, in the Schwarzschild-type gauge the perturbed metric takes the form $\delta g^{\mu\nu}=\delta^{\mu\nu}\delta f$.

With the specifications mentioned above, we now proceed to derive equation \eqref{eq:quadratic_action_SRG}. To accomplish this, we examine the integrals $I_i$ defined in Appendix \ref{sec:derivation_quadratic} that make up the quadratic action. By setting $\bar\Phi=ar$ and $k=1/2$ in equation \eqref{eq:I_1_3} and carrying out an integration by parts we obtain
\begin{align}\label{eq:I_1_SRG}
I^{\text{SRG}}_1=4\int\text{d}^2x\ \partial^\mu\delta\Phi\partial_\mu\delta\Phi\ ,
\end{align}
where $\bar\Box:=\eta^{\mu\nu}\partial_\mu\partial_\nu$ is the d'Alembertian in flat space-time. In order to compute the remaining $I^{\text{SRG}}_i$ integrals, we substitute in $\delta g^{\mu\nu}=\delta^{\mu\nu}\delta f$. Using the fact that $\delta^{\mu\nu}\eta_{\mu\nu}=0$, the remaining $I^{\text{SRG}}_i$ integrals simplify considerably since the contributions involving contractions $\delta g^{\mu\nu}\eta_{\mu\nu}$ will vanish. For $I^{\text{SRG}}_2$ and $I^{\text{SRG}}_3$ we find
\begin{align}\label{eq:I2+I3_SRG_1}
I^{\text{SRG}}_2+I^{\text{SRG}}_3&=4\int\text{d}^2x\ \delta^{\mu\nu}\delta f\Big[2\partial_\mu\bar\Phi\partial_\nu\delta\Phi\nonumber\\
&-\partial_\mu\partial_\nu\left(\bar\Phi\delta\Phi\right)\Big]\ .
\end{align}
Since the background dilaton is $\bar\Phi=ar$, we have $\partial_\mu\partial_\nu\bar\Phi=0$ and thus \eqref{eq:I2+I3_SRG_1} becomes
\begin{align}\label{eq:I2+I3_SRG_2}
I^{\text{SRG}}_2+I^{\text{SRG}}_3=-4\int\text{d}^2x\ \delta^{\mu\nu}\delta f\bar\Phi\partial_\mu\partial_\nu\delta\Phi\ .
\end{align}
We now turn our attention to $I^{\text{SRG}}_4$. From equation \eqref{eq:I_4_3} we have
\begin{align}
&I^{\text{SRG}}_4=a^2\int\text{d}^2x\ \delta f\delta^{\mu\nu}\bigg\{\eta_{\mu\alpha}\Big[2\partial_\beta r\partial_\nu r-\partial_\nu\partial_\beta r^2\Big]\nonumber\\
&+\frac12\eta_{\mu\alpha}\eta_{\nu\beta}\partial^\sigma r^2\partial_\sigma-\eta_{\mu\beta}\partial_\alpha r^2\partial_\nu\bigg\}\delta f\delta^{\alpha\beta}\ ,
\end{align}
where we have once again used $\delta^{\mu\nu}\eta_{\mu\nu}=0$. By evaluating the derivatives, the contribution contained in the square brackets vanishes and we are left with
\begin{align}
I^{\text{SRG}}_4=a^2\int\text{d}^2x\ r\delta f\Big(\eta_{\mu\nu}\eta^{\mu\nu}-2\Big)\partial_r\delta f\ .
\end{align}
It follows that, since the trace of the Minkowski metric in two space-time dimensions is $\eta^\mu_{\ \mu}=2$, the integral above vanishes and we have
\begin{align}\label{eq:I_4_SRG}
I^{\text{SRG}}_4=0\ .
\end{align}
Upon the substitution of equations \eqref{eq:I_1_SRG}, \eqref{eq:I2+I3_SRG_2} and \eqref{eq:I_4_SRG} into \eqref{eq:quadratic_action_I_i} we obtain the desired result \eqref{eq:quadratic_action_SRG}.
\section{Derivation of equation \texorpdfstring{\eqref{eq:CGHS_non_local_dilaton_solution}}{}}\label{section:derive_CGHS_non_local_dilaton}

In this section, we evaluate the integral appearing in equation \eqref{eq:CGHS_inverse_Fourier_space_solution_non_local} and obtain the non-local modification to the perturbed dilaton field \eqref{eq:CGHS_non_local_dilaton_solution}. We begin by writing equation \eqref{eq:CGHS_inverse_Fourier_space_solution_non_local} as
\begin{align}\label{eq:CGHS_solve_non_local_dilaton}
\delta\Phi=&\frac{M{\text e}^{-\lambda r}}{4\pi i\lambda}\int^\infty_{-\infty}\text{d}k\ {\text e}^{-\ell^2k^2}\nonumber\\
&\times\bigg[i\int^{r-b}_0\text{d}u\ {\text e}^{iu(k-i\lambda)}+\frac{1}{k-i\lambda}\bigg]\ .
\end{align}
We first wish to compute the second term on the right-hand-side of the above expression. In order to accomplish this, we need to evaluate the integral
\begin{align}
L_1:=\int\text{d}k\ \frac{{\text e}^{-\ell^2k^2}}{k-i\lambda}\ .
\end{align}
The real part of the integral given above is
\begin{align}
\text{Re}\left\{L_1\right\}&=\int\text{d}k\ \frac{{\text e}^{-\ell^2k^2}k}{k^2+\lambda^2}\nonumber\\
&=0\ ,
\end{align}
which vanishes since the integrand is an odd function of $k$. This implies that the integral $L_1$ is purely imaginary. That is,
\begin{align}
L_1=i\text{Im}\left\{L_1\right\}=i\lambda\int\text{d}k\ \frac{{\text e}^{-\ell^2k^2}}{k^2+\lambda^2}\ .
\end{align}
The last expression can be written as
\begin{align}
L_1=-i\lambda {\text e}^{\ell^2\lambda^2}\int^\infty_{-\infty}\text{d}k\int^{\ell^2}_\infty\text{d}u\ {\text e}^{-u\left(k^2+\lambda^2\right)}\ .
\end{align}
By evaluating the integral over $k$, one finds
\begin{align}
L_1=-i\lambda\sqrt\pi {\text e}^{\ell^2\lambda^2}\int^{\ell^2}_\infty\text{d}u\ \frac{{\text e}^{-u\lambda^2}}{\sqrt u}\ .
\end{align}
Performing a change of variables with $u=v^2/\lambda^2$ gives
\begin{align}\label{eq:L1_integral_final}
L_1=2i {\text e}^{\ell^2\lambda^2}\sqrt\pi\int_{\ell\lambda}^\infty\text{d}v\ {\text e}^{-v^2}\ .
\end{align}
The integral is nothing more than $\frac{\sqrt\pi}2\text{Erfc}(\ell\lambda)$ where $\text{Erfc}$ is the complementary error function. We therefore have
\begin{align}
L_1=i\pi {\text e}^{\ell^2\lambda^2}\text{Erfc}(\ell\lambda)\ .
\end{align}
We now wish to compute the first term on the right-hand-side of equation \eqref{eq:CGHS_solve_non_local_dilaton}. We therefore turn our attention to evaluating the integral
\begin{align}
L_2:&=i\int^\infty_{-\infty}\text{d}k\int^{r-b}_0\text{d}u\ {\text e}^{-\ell^2k^2+iku+\lambda u}\nonumber\\
&=i\int^{r-b}_0\text{d}u\int^\infty_{-\infty}\text{d}k\ {\text e}^{-\ell^2\left(k-\frac{iu}{2\ell^2}\right)^2-\frac{u^2}{4\ell^2}+\lambda u}\ .
\end{align}
Evaluating the integral over $k$ gives us
\begin{align}
L_2=i\frac{\sqrt\pi}{\ell}\int^{r-b}_0\text{d}u\ {\text e}^{-\frac{1}{4\ell^2}\left(u-2\lambda\ell^2\right)^2+\lambda^2\ell^2}\ .
\end{align}
By performing the change of variables $v=(2\lambda\ell^2-u)/2\ell$ we obtain
\begin{align}\label{eq:L2_integral_final}
L_2=2i\sqrt\pi {\text e}^{\ell^2\lambda^2}\int^{\lambda\ell}_{\lambda\ell-\frac{r-b}{2\ell}}\text{d}v\ {\text e}^{-v^2}\ .
\end{align}
By adding equations \eqref{eq:L1_integral_final} and \eqref{eq:L2_integral_final} we find
\begin{align}\label{eq:L_1+L_2}
L_1+L_2=i\pi {\text e}^{\ell^2\lambda^2}\text{Erfc}\left(\ell\lambda-\frac{r-b}{2\ell}\right)\ .
\end{align}
Substituting equation \eqref{eq:L_1+L_2} into \eqref{eq:CGHS_solve_non_local_dilaton} gives the desired result \eqref{eq:CGHS_non_local_dilaton_solution}.
\section*{Acknowledgements}
UKBV acknowledges financial support from the University of Cape Town Postgraduate Funding Office through the Mathematics and Applied Mathematics Departmental Scholarship, the Doctoral Research Scholarship and the Vice Chancellor Research Scholarship. AdlCD acknowledges support from NRF grants no.120390, reference:BSFP190416431035; no.120396, reference:CSRP190405427545; PID2019-108655GB-I00 and COOPB204064, I-COOP+2019, MICINN Spain. IK and AM are supported by Netherlands Organization for Scientific Research (NWO) grant no. 680-91-119. FJMT acknowledges financial support from ``Fundaci\'on Ram\'on Areces''; NRF grants no. 120390, reference: BSFP190416431035; no. 120396, reference: CSRP190405427545; no. 101775, reference: SFH150727131568, the Research Council of Norway, and the European Regional Development Fund through the Center of Excellence TK133 ``The Dark Side of the Universe''.
\bibliographystyle{apsrev4-1}
\end{document}